\documentclass[sigconf,final]{acmart}
\AtBeginDocument{%
  }

\setcopyright{acmlicensed}
\copyrightyear{2025}
\acmYear{2025}
\setcopyright{acmlicensed}\acmConference[KDD '25]{Proceedings of the 31st ACM SIGKDD Conference on Knowledge Discovery and Data Mining V.2}{August 3--7, 2025}{Toronto, ON, Canada}
\acmBooktitle{Proceedings of the 31st ACM SIGKDD Conference on Knowledge Discovery and Data Mining V.2 (KDD '25), August 3--7, 2025, Toronto, ON, Canada}
\acmDOI{10.1145/3711896.3736981}
\acmISBN{979-8-4007-1454-2/2025/08}

\usepackage{amsmath}
\usepackage{amsfonts}
\usepackage{tabularx}  
\usepackage{graphicx}  
\usepackage{array}  
\usepackage{multirow} 
\usepackage{xspace}
\usepackage{booktabs}
\usepackage{tcolorbox}
\usepackage{listings}%
\usepackage{subfig}

\newcommand{\ie}{\emph{i.e.,}\xspace}
\newcommand{\aka}{\emph{a.k.a.,}\xspace}
\newcommand{\eg}{\emph{e.g.,}\xspace}

\newcommand{\ignore}[1]{}

\definecolor{promptboxcolor}{RGB}{253, 240, 220} 

\newcommand{\fullmodel}{\textbf{G}enerative \textbf{N}ext \allowbreak \textbf{P}OI \allowbreak \textbf{R}ecommendation \allowbreak with \allowbreak \textbf{S}emantic \textbf{ID}\xspace}
\newcommand{\model}{GNPR-SID\xspace}

\begin{document}

\title{Generative Next POI Recommendation with Semantic ID}

\author{Dongsheng Wang}
\affiliation{%
  \institution{University of Electronic Science and Technology of China}
  \city{Chengdu}
  \country{China}
}
\email{202411081626@std.uestc.edu.cn}

 \author{Yuxi Huang}
\affiliation{%
  \institution{University of Electronic Science and Technology of China}
  \city{Chengdu}
  \country{China}
}
\email{yuxi.h.cs@gmail.com}

\author{Shen Gao}
\authornote{Shen Gao and Shuo Shang are corresponding authors.}
\affiliation{%
  \institution{University of Electronic Science and Technology of China}
  \city{Chengdu}
  \country{China}
}
\email{shengao@uestc.edu.cn}

\author{Yifan Wang}
\affiliation{%
  \institution{University of Electronic Science and Technology of China}
  \city{Chengdu}
  \country{China}
}
\email{202422081324@std.uestc.edu.cn}

\author{Chengrui Huang}
\affiliation{%
  \institution{University of Electronic Science and Technology of China}
  \city{Chengdu}
  \country{China}
}
\email{2021080101014@std.uestc.edu.cn}

\author{Shuo Shang}
\authornotemark[1]
\affiliation{%
  \institution{University of Electronic Science and Technology of China}
  \city{Chengdu}
  \country{China}
}
\email{jedi.shang@gmail.com}

\renewcommand{\shortauthors}{Dongsheng Wang et al.}

\begin{abstract}
  Point-of-interest (POI) recommendation systems aim to predict the next destinations of user based on their preferences and historical check-ins.
  Existing generative POI recommendation methods usually employ random numeric IDs for POIs, limiting the ability to model semantic relationships between similar locations. 
  In this paper, we propose \fullmodel (\model), an LLM-based POI recommendation model with a novel semantic POI ID (SID) representation method that enhances the semantic understanding of POI modeling. 
  There are two key components in our \model: (1) a \textbf{Semantic ID Construction} module that generates semantically rich POI IDs based on semantic and collaborative features, and (2) a \textbf{Generative POI Recommendation} module that fine-tunes LLMs to predict the next POI using these semantic IDs. 
  By incorporating user interaction patterns and POI semantic features into the semantic ID generation, our method improves the recommendation accuracy and generalization of the model. 
  To construct semantically related SIDs, we propose a POI quantization method based on residual quantized variational autoencoder, which maps POIs into a discrete semantic space. 
  We also propose a diversity loss to ensure that SIDs are uniformly distributed across the semantic space.
  Extensive experiments on three benchmark datasets demonstrate that \model substantially outperforms state-of-the-art methods, achieving up to 16\% improvement in recommendation accuracy\footnote{Code is available at \url{https://github.com/wds1996/GNPR-SID}}.
  
\end{abstract}



\maketitle

\section{Introduction}
Location-based social networks, such as Foursquare and Google Maps, are being used more and more frequently in modern life, with a large number of users sharing their check-in records on these platforms via mobile devices, such as once-trip or once-restaurant. 
Thus, point-of-interest (POI) recommendation~\cite{islam2022survey,liu2017experimental,sanchez2022point} became a popular research task, which aims to provide location services to users based on their preferences and context information. 
The next POI recommendation~\cite{cheng2013you,lian2020geography,zhao2020go}, is a variant of POI recommendation tasks,  focuses on the user's historical check-in trajectory, incorporating sequential transitions and spatial-temporal context information to predict the next visit.

Recently, with the success of Large Language Models (LLMs) in various domains~\cite{brown2020language,bai2023qwen,dubey2024llama}, LLM-based generative recommendation has emerged as a new paradigm for next POI recommendation~\cite{li2024large,feng2024move,wang2024large,wongso2024genup}, which directly generates the next POI IDs for a given user query. 
Compared to traditional POI recommendation methods, these models can be trained end-to-end, which avoids using the embedding vector for each POI and instead allows the model to directly generate the POI ID. 
This improves the generalization ability of model and reduces the space for storing the vectors of each POI.

\begin{figure}
    \centering
    \includegraphics[width=0.90\linewidth]{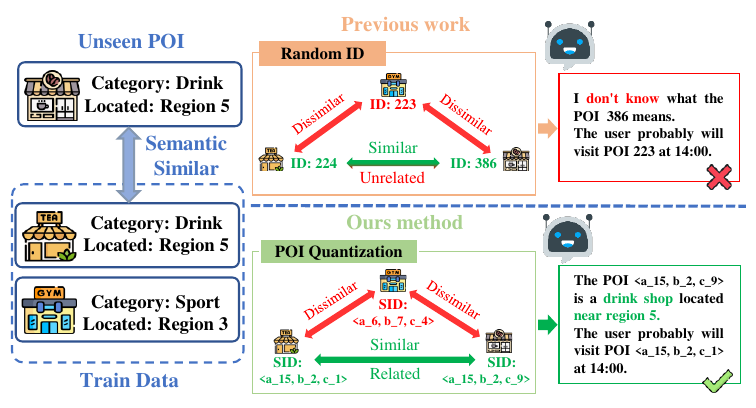}
    \caption{Comparison of random ID and semantic ID: Semantic ID can captures semantic associations between similar POIs which can facilitate LLM to model the user preference.}
    \label{fig:comparison}
\end{figure}

Existing LLM-based POI recommendation methods~\cite{li2024large,wongso2024genup} rely predominantly on Random ID (RID), where each POI is treated as a random unique numeric representation, and user historical interactions are modeled as sequences of these RIDs. 
Traditional methods based on POI embedding vectors can store rich semantic information about POIs within the embedding vectors. 
However, existing LLM-based methods only use the randomly generated RID as input, and these random numbers do not capture any semantic information. 
As a result, existing LLM-based methods struggle to effectively model POI semantic information. 
It is also evident that these methods also exhibit poor generalization ability for POI information outside the dataset.

Therefore, designing an effective POI ID representation method is crucial for improving LLM-based POI recommendation models. 
A well-designed POI ID representation method should have the following characteristics: 
(1) The POI ID should reflect the semantic information of the POI, with similar POIs having similar IDs; 
(2) It should represent a vast number of POIs using a smaller ID space, without causing ID conflicts.
For example, in a nearby neighborhood, there is a coffee shop and a tea shop. 
As adjacent stores sell similar beverages, both designed to refresh customers, these two shops should have similar IDs. 
We show a comparison between existing LLM-based POI recommendation methods and our proposed \model in Figure~\ref{fig:comparison}.
For instance, if the coffee shop's ID is <a\_15>< b\_2>< c\_9>, a good ID for the tea shop might be <a\_15>< b\_2>< c\_1>. 
This ID representation reflects the similarity between the two POIs, which not only helps the LLM better understand the semantics of the POIs but also allows for better generalization when a new bubble tea shop is added. 
By assigning an ID like <a\_15>< b\_2>< c\_x> to the bubble tea shop, the system can help the LLM generalize to the newly introduced POI more effectively.
However, the existing methods assign a random ID (\eg 386) for the newly added POI which cannot reflect any semantic similarity between the other two shops (\eg 223 and 224).

Therefore, in this paper, we propose \fullmodel (\model), which introduces an LLM-based POI recommendation model with a novel semantic POI ID representation module that enhances the semantic understanding of POI modeling.
There are two main modules in \model: 
(i) \textbf{Semantic ID construction module}, which can generate a semantic ID for each POI based on common POI features (\eg location, category, etc.), where the semantic ID consists of multiple numbers and POIs with similar semantics share similar ID prefixes. 
(ii) \textbf{Generative POI recommendation module}, which uses the semantic POI ID as input and fine-tunes the LLM to predict the next POI by incorporating the knowledge and reasoning ability within the LLM.

Since POI features are typically represented by numerical data, which are challenging for LLMs primarily designed for text understanding, we propose a \textbf{POI semantic representation} method. 
This method transforms complex numerical features into semantic representations that are more interpretable for LLMs. 
Furthermore, our approach not only captures the intrinsic semantic information of POIs but also integrates collaborative filtering signals and user visiting patterns into the POI semantic ID, enriching the representation with both contextual and behavioral insights.
Next, we propose a \textbf{hierarchical POI quantization} method based on residual quantized variational autoencoder.
This process maps POI embeddings into a discrete semantic space, where each POI is represented by a sequence of codeword indices derived from multi-layer codebooks. 
This quantization ensures that semantically similar POIs share similar SIDs, preserving their relationships in a structured and compact form.
To achieve this, we propose the diversity loss, which promotes the distinctiveness of codewords within each codebook layer.

Finally, we use the POI SID representation as input to the LLM, and end-to-end to train the model to directly predict the next POI based on the user historical check-in data. 
Compared to existing LLM-based next POI recommendation methods, our approach helps the model comprehensively understand the semantics of each POI and the deeper relationships between POIs. 
By leveraging the rich commonsense knowledge and reasoning capabilities inherent in LLMs, our method enables a better understanding of user intent, leading to more accurate next POI recommendations.
To demonstrate the effectiveness of \model, we conduct extensive experiments on three real-world benchmark datasets: Foursquare-NYC~\cite{yang2014modeling}, Foursquare-TKY~\cite{yang2014modeling}, and GowallaCA~\cite{cho2011friendship}. 
Experimental results demonstrate that \model consistently outperforms state-of-the-art approaches, with accuracy improvements reaching up to 16\% on these benchmark datasets. 
Furthermore, to validate that the generated SIDs effectively capture the semantic similarity between POIs, we conduct semantic similarity analysis and visualization experiments on the SIDs. 
The experiments demonstrate that semantically similar POIs share more common SID prefixes.

\noindent Our contributions are summarized as follows:

\noindent $\bullet$ We first introduce the \fullmodel (\model), which fine-tunes the LLM to comprehensively understand the spatial-temporal semantics in the POI recommendation task.

\noindent $\bullet$ We propose the semantic encoding of POI metadata and then propose a semantic ID construction method that maps POI into semantic ID by POI quantization.

\noindent $\bullet$ We train an LLM to directly generate the next POI SID by incorporating the knowledge and reasoning ability of an LLM.

\noindent $\bullet$ Extensive experiments on three benchmark datasets demonstrate that \model outperforms state-of-the-art methods.

\section{Related Work}

Early next POI recommendation methods are often framed as a sequential recommendation task, leveraging feature engineering such as personalized Markov chains~\cite{cheng2013you,he2016inferring}, but suffered from limited scalability and dependency on domain expertise. 
With the rise of deep learning, RNN-based models~\cite{liu2016predicting,kong2018hst,sun2020go} were widely adopted. 
For example, \citet{zhao2020go} propose STGN a spatio-temporal gated network, to capture personalized sequential patterns, addressing user long and short-term preferences. 
Driven by the flexibility of attention mechanisms, Transformer-based models~\cite{wang2022spatial,zhang2022next,duan2023clsprec,wu2020personalized} are surpassing RNNs in next POI recommendation.
For example, STAN~\cite{luo2021stan} proposes a spatial-temporal attention mechanism to capture spatial-temporal relevance within POI trajectories. 
In addition, graph neural networks gained traction for modeling user-POI interactions~\cite{wang2023adaptive,wang2022learning,lim2020stp,lim2022hierarchical,yang2022getnext}. 
For example, STHGCN~\cite{yan2023spatio} uses a spatial-temporal hypergraph convolutional network that captures trajectory-level information by learning from both user history and collaborative trajectories. 

Recently, LLMs have demonstrated strong capabilities in handling spatial-temporal data and commonsense knowledge, achieving significant success in retrieval tasks~\cite{sun2024learning,wang2022neural} and recommendation tasks~\cite{rajput2023recommender,wang2024enhanced,wang2024learnable}, making them a promising solution for next POI recommendation. 
LLM-Mob~\cite{wang2023would} introduce in-context learning to enhance next POI recommendation using historical and contextual trajectories. 
NextLocLLM~\cite{liu2024nextlocllm} uses spatial coordinates to enhance its understanding of spatial relationships between locations, transforming ID prediction into coordinate prediction, and improving transferability and generalization in different datasets.
LLM4POI~\cite{li2024large} performs supervised fine-tuning of the LLM to transform the next POI recommendation task into a question-answering task, and it achieves the stats-of-the-art performance in several next POI recommendation benchmark datasets.
These models rely solely on randomly initialized ID representations or a combination of random numeric IDs with text prompts for POI trajectory input, failing to capture semantic associations between POIs and demonstrating limited cross-dataset generalization capabilities.
However, the semantic association between POIs is not only reflected in spatial coordinates.

To address the challenge of incorporating non-textual knowledge into LLMs, multimodal large language models (MLLMs)~\cite{Bai2023QwenVL,Shu2023MLLM} provide a perspective. OneLLM~\cite{Han2024OneLLM} maps signals from any modality into embeddings using a frozen unified encoder and an MOE-based Universal Projection Module, then inputs them into LLMs for downstream tasks. However, POI representation requires modeling temporal sequences and spatial location information, most existing MLLMs cannot effectively capture POI semantic information. Therefore, exploring the application of MLLM methods to POI recommendation is also a promising research direction worth investigating.

\begin{figure*}[h]
    \centering
    \includegraphics[width=1\linewidth]{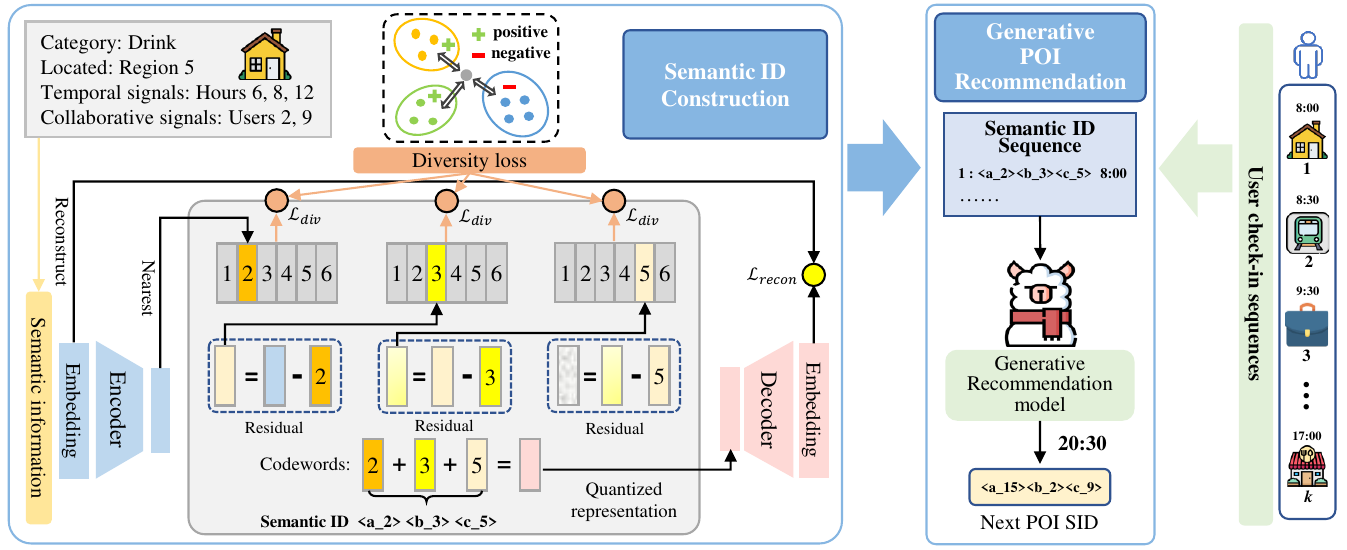}
    \caption{The overview of the \fullmodel (\model). It consists of two modules: (1) \textit{Semantic POI ID Construction} module maps POI to a semantic ID (SID) using the codebook quantization and (2) \textit{Generative POI Recommendation} module fine-tunes an LLM to use historical POI SID sequence as input and generate the next POI SID.}
    \label{fig:overall-arch}
\end{figure*}

\section{Problem Definition} 
\label{definition}

Let $ U = \{ u_1, u_2, \dots, u_M \} $ and $ P = \{ p_1, p_2, \dots, p_N \} $ represent the set of $M$ users and $N$ POIs, respectively.
Each POI is characterized by a tuple $p_i = (c, lon, lat, u_{p_i}, t_{p_i})$, where $c$ represents the category (\eg Home, Subway and Food Shop), $lon$ and $lat$ denotes the geographical coordinates (longitude and latitude), $u_{p_i}$ refers to the set of check-in users, and $t_{p_i}$ indicates the set of timestamps when users visited. 
For a user $ u \in U $, we can obtain the recent visited POIs $ P_u = \{p_1, p_2, \ldots, p_i\} $, and the corresponding check-in timestamps for each POI $ T_u = \{t_1, t_2, \ldots, t_i\} $.
In the next POI recommendation task, we use the recent check-in record $P_u$ and $T_u$ as input to predict the next POI $p_{i+1}$ that the user will visit at time $t_{i+1}$.

\section{Methodology}
\label{sec:meth}

The overall framework of \model is presented in Figure~\ref{fig:overall-arch}. 
The framework consists of two main modules: (1) the \textbf{semantic ID construction} module maps each POI to a semantic ID (SID) using the codebook quantization, and (2) the \textbf{generative POI recommendation} module fine-tunes an LLM to generate the next POI which leverages the SID as the representation of POI.

\subsection{Semantic ID Construction}
\label{sec:SIDs}

To map the POI features to the SID, we first construct a semantic codebook, which is an autoencoder framework to encode the POI dense representation into a sparse space and then reconstruct the POI dense representation.
A well-designed SID representation enhances the performance of subsequent LLM-based recommendation models by capturing semantic correlations between POIs. 
Specifically, an effective SID should exhibit three key characteristics:
(i) \textbf{Semantic Richness}: It should encapsulate significant semantic information of the POI.
(ii) \textbf{Semantic Similarity}: POIs with similar semantics should possess similar SIDs.
(iii) \textbf{Uniqueness}: Distinct POIs should have unique SIDs to avoid conflicts.

\subsubsection{POI Semantic Representation}

The first challenge in mapping POIs to semantic IDs is effectively capturing and representing their semantic information. 
Typically, a POI is described by its category and discrete latitude-longitude coordinates. 
However, these attributes alone are insufficient to capture semantic associations between POIs, and discrete coordinates are suboptimal for model learning. 
To address this limitation, we propose a comprehensive semantic representation method for POI data. 
Specifically, we extract four key features to represent a POI: 
(1) basic category information, 
(2) location data, 
(3) temporal signals, and 
(4) user collaborative signals. 
This multi-faceted representation enriches the semantic understanding of POIs, enabling more accurate and meaningful mappings.

\paragraph{Spatial Features}

To capture spatial features (\eg, longitude and latitude), we utilize Google Maps' Plus Codes\footnote{\url{https://maps.google.com/pluscodes/}} to convert latitude-longitude coordinates into region-based codes, denoted as $r$, which represent unique numeric region IDs. 
The plus code defines regions by recursively dividing areas using grids of varying sizes and adjusting the code length.
Compared to directly using geographic coordinates as spatial features, these region-based codes $r$ ensure that nearby POIs are assigned the same Plus Code, allowing the model to share information among spatially adjacent POIs. 
Such spatial grouping not only simplifies the representation of geographic relationships but also enhances the model to capture regional patterns and dependencies. 

\paragraph{Temporal Signals}

To extract temporal features, we partition a day into 24-time slots and map the historical visit times of each POI to these slots. 
Time slots with the highest visitation frequency are regarded as the potential operating hours of the POI, represented by $t$.
The top 10 time slots with the highest visitation frequency are used in our experiments.
Compared to directly using the raw timestamps, the time slots $t$ provide a more structured and interpretable representation of temporal patterns, which can effectively capture the temporal dynamics of POI visits.

\paragraph{Collaborative Signals}

Additionally, to model user collaborative signals, we analyze the 10 most frequent visitors for each POI and analyze their co-visitation patterns.
This enables the model to capture latent connections between POIs through shared user behavior.
The collaborative information is represented by $c_u$, which captures user collaboration signals based on the set of check-in users with shared interaction histories.
This approach enriches the representation of POIs by incorporating both temporal dynamics and collaborative user behavior.

Finally, we construct one-hot vectors for each of the aforementioned features.
The semantic representation of a POI, denoted as vector $ p_e $, is defined as:
\begin{equation}
\label{eq:poi_representation}
    p_e = \text{concat}(c, r, t, c_u),
\end{equation}
where $ c $ represents the POI category, $ r $ denotes the region, $ t $ corresponds to the operating time slots, and $ c_u $ captures the user collaboration signals based on shared interaction histories. 
Here, $ \text{concat}(\cdot) $ refers to the concatenation operation. 
This unified representation $ p_e $ effectively integrates multiple semantic aspects of a POI, enabling a comprehensive and structured input for downstream tasks.

\subsubsection{POI Quantization} 

After constructing the semantic vector representation $ p_e $ for each POI, the next task is to generate an ID for each POI according to the representation $ p_e $, which can represent the correlation between semantically similar POIs. 
Specifically, semantically similar POIs should share as many prefix codes as possible. 
For example, the SID <a\_15><b\_2><c\_1> should be more similar to the POI with SID <a\_15><b\_2><c\_9> than to a POI with SID <a\_15><b\_12><c\_2>. 
This prefix-based similarity mechanism enhances the robustness and interpretability of the SID structure, facilitating more accurate and meaningful recommendations.

Thus, we propose a novel SID quantization approach for POIs. 
This method ensures that the generated SIDs can represent semantic relationships among POIs, thereby improving the robustness and relevance of recommendations which can be easily adapted to the LLM-based generative model. 
Specifically, we adopt the Residual Quantized Variational AutoEncoder (RQVAE)~\cite{zeghidour2021soundstream}, a multi-level vector quantizer that leverages residual quantization, to map POI embeddings into SIDs. 
The RQVAE framework comprises three key modules: an encoder, a residual quantization module, and a decoder. 
This architecture enables hierarchical and fine-grained quantization, effectively capturing the semantic structure of POIs.

The encoder extracts a deep representation from the input POI vector, which is then used for subsequent quantization:
\begin{equation}
    z_e = \text{Encoder}(p_e),
\end{equation}
where $ p_e $ is the input POI vector (in Equation~\ref{eq:poi_representation}), and $ z_e $ represents the hidden representation of $ p_e $ obtained through the encoder. 
The encoder is implemented using fully connected layers, which effectively capture the complex semantic features of the POI. 
This hidden representation $ z_e $ serves as the foundation for the subsequent quantization process.

First, we define a hierarchical codebook space $B$ consisting of $L$ layers, where each layer contains $K$ codeword vectors. 
Each codeword vector is represented in a $d$-dimensional space:
\begin{equation}
    B^{(l)} = \{e_1, e_2, \dots, e_K\},
\end{equation}
where $B^{(l)}$ represents the codebook space of the $l$-th layer, and $e_k \in \mathbb{R}^{K \times d}$ denotes the $k$-th codeword vector in the layer. 
This hierarchical structure enables multi-level quantization, allowing the model to capture fine-grained semantic relationships among POIs.
Residual quantization comprises $L$ layers of quantizers (\aka $L$-layer codebook spaces). 
Each quantizer maps the input vector $r^{(l)}$ to the closest discrete representation (codeword vector) in its corresponding layer:
\begin{equation}
{e_k}^{(l)} = \arg\min_{e \in B^{(l)}} \left\|{r}^{(l)}-e \right\|, 
\end{equation}
where the  $e_k^{(l)}$  is the quantized codeword at level  $l$  and the  ${r}^{(l)}$ denotes the from the residual quantizer of the $(l-1)$-th layer. 
These codewords serve as cluster nodes, enabling the indexing of POIs as sequences of codewords with hierarchical information. This cluster process is repeated recursively $L$ times to get a tuple of  $L$ codewords that represent the Semantic ID. This recursive approach approximates the input from a coarse-to-ﬁne granularity. Note that distinct codebook spaces are used for each layer, to attain a more refined representations of indices in a coarse-to-fine manner. 
The input to each quantizer layer is the residual from the previous layer of quantization:
\begin{equation}
{r}^{(l)} = {r}^{(l-1)} - {e_k}^{(l-1)}.
\end{equation}
For the first layer ($l=0$), the input residual is initialized as ${r}^{(0)} = z_e$, since $z_e$ serves as the input to the first layer. 
This hierarchical residual quantization process ensures fine-grained representation and preserves semantic relationships across layers.
After recursive quantization for a POI, the quantized vector $ \hat{z_e} $ is obtained by summing the codeword vectors from each layer:
\begin{equation}
    \hat{z_e} = \sum_{l=1}^{L} {e_k}^{(l)},
\end{equation}
where $k \in \{1, \dots, K\}$, and ${e_k}^{(l)}$ represents the codeword vector obtained from the $l$-th layer quantization. 
To maintain the semantic correlation between POI and SID, we employ the decoder to reconstruct POI vector vector ${z_e}$ from the quantized vector $ \hat{z_e} $:
\begin{equation}
    \hat{p_e} = \text{Decoder}(\hat{z_e}),
\end{equation}
where the decoder is also implemented by the DNN.

Based on the hierarchical codebook spaces we constructed, we generate the SID by forming a tuple of codeword indices, where each index corresponds to a codeword vector from a distinct layer of the codebook space $B^{(l)}$:
\begin{equation}\label{equ:build-sid}
\text{SID} = [\text{index}({e}^{(1)}), \text{index}({e}^{(2)}), \dots, \text{index}({e}^{(L)})],
\end{equation}
where ${e}^{(l)} \in B^{(l)}$ represents the codeword vector from the $l$-th layer, and $\text{index}(\cdot)$ retrieves the position of the codeword vector within its respective layer. 
For instance, a SID of the form <a\_15><b\_2><c\_1>, derived from a three-layer codebook structure,   indicates that the indices <a\_15>, <b\_2>, and <c\_1> correspond to the codeword vectors ${e_{15}}^{(1)}$, ${e_2}^{(2)}$, and ${e_1}^{(3)}$ from the first, second, and third layers, respectively. 
This structured representation ensures that the SID captures the hierarchical semantic relationships among POIs, facilitating efficient and interpretable recommendations.

To further ensure the uniqueness of SIDs, we address any remaining cases of SID collisions by introducing an additional dimension to the SIDs. 
Specifically, if multiple POIs are mapped to the same SID, we append a unique identifier to each SID to distinguish them. 
For example, if two POIs are both mapped to SID <a\_15><b\_2><c\_9>, they will be represented as <a\_15><b\_2><c\_9><d\_0> and <a\_15> <b\_2><c\_9><d\_1>, respectively. 
This approach guarantees that each POI has a distinct SID, even in cases where semantic collisions occur in the initial quantization process. 
By incorporating this additional dimension, we enhance the robustness and uniqueness of the SID representation, ensuring that the model can accurately differentiate between POIs with similar semantic features.

\subsubsection{Model Optimization}

\paragraph{Reconstruction Loss}

To optimize the model parameters, we employ the reconstruction loss, which quantifies the distance between the original embedding ${p_e}$ and the reconstructed embedding $\hat{p_e}$:
\begin{equation}\label{equ:reconstruction-loss}
\mathcal{L}_{\text{recon}} = \| p_e - \hat{p_e} \|^2 .
\end{equation}

\paragraph{Quantization Loss}

Additionally, we introduce a quantization loss $\mathcal{L}_{\text{quant}}$ to evaluate the discrepancy between the residuals of the previous layer and the codeword vectors of the current layer. 
This loss reflects the gap between the input and its quantized representation, ensuring minimal information loss during the quantization process. 
The quantization loss $\mathcal{L}_{\text{quant}}$ consists of two components: (1) the quantization loss, which measures the distance between the current input vector and the codeword vector, and (2) the commitment loss, which encourages the input vector to align more closely with the codeword vector through an additional constraint term:
\begin{equation}\label{equ:quant-loss}
    \mathcal{L}_{\text{quant}} = \sum_{l=1}^{L} \| \text{sg}[\mathbf{r}^{(l-1)}] - \mathbf{e_k}^{(l)} \|^2 + \beta \| \mathbf{r}^{(l-1)} - \text{sg}[\mathbf{e_k}^{(l)}] \|^2 ,
\end{equation}
where $\text{sg}$ denotes the stop-gradient operation~\cite{van2017neural}, and $\beta$ is a hyperparameter that controls the strength of the commitment loss. 
By optimizing $\mathcal{L}_{\text{quant}}$, we ensure that the quantized representation closely approximates the original input, preserving semantic fidelity throughout the quantization process.

\paragraph{Diversity Loss} 

To ensure the uniqueness of SIDs and avoid semantic collisions where multiple POIs map to the same SID, it is crucial to impose constraints during the POI quantization process. 
Without such constraints, relying solely on the reconstruction loss could lead to a large number of POIs being quantized into a small subset of semantic spaces, compromising the distinctiveness of SIDs. 
Therefore, we introduce a diversity loss to promote the diversity of codewords within the codebook. 
This loss encourages the model to capture distinct semantic information and ensures more efficient utilization of the discrete quantization space. 
We define the diversity loss from two complementary perspectives: codeword utilization constraint and intra-codeword compactness constraint.
The diversity loss is defined as:
\begin{equation}
\mathcal{L}_{\text{div}} = \mathcal{L}_{\text{utilize}} + \sum_{i=1}^{K}\mathcal{L}_{\text{compactness}(k_i)},
\label{equ:diversity-loss}
\end{equation}
Specifically, we aim to encourage POIs within a batch to be evenly mapped across different codeword spaces, thereby maximizing the utilization of the codebook. 
\begin{equation}
\mathcal{L}_{\text{utilize}} = \frac{1}{N}\sum_{i=1}^{K}\| count_i-\frac{N}{K} \|,
\end{equation}
where $N$ is the total number of data, K is the number of codewords (reflected in Fig\ref{fig:overall-arch} is the number of clusters), $count_i$ denotes the number of vectors now existing in the $i$-th codeword space.

However, optimizing solely for codeword utilization may reduce semantic coherence, forcing semantically dissimilar POIs into the same cluster for distributional balance. 
To address this, we add a constraint enforcing semantic compactness within each codeword, ensuring vectors mapped to the same codeword are close in embedding space. 
\begin{equation}
\mathcal{L}_{\text{compactness}} = \frac{1}{k{(k-1)}} \sum_{i \neq j}^{k} \| e_i - e_j \|^2,
\end{equation}
where $\| e_i - e_j \|^2$ measures the distance between codeword vectors $i$ and $j$ in current codeword space, k is the total number of vectors in current codeword space. 
This dual-constraint design maintains high codebook utilization while preserving semantic consistency.  
By optimizing the diversity loss, the model maximizes codebook utilization while minimizing intra-codeword variance, thereby reducing semantic collisions and enhancing the discriminative power and uniqueness of the generated SIDs.
 
\paragraph{Total Loss} 

To optimize the model, the total loss is defined as a weighted combination of the reconstruction loss, quantization loss, and diversity loss:
\begin{equation}
     \mathcal{L}_{\text{total}} = \mathcal{L}_{\text{recon}} + \mu \mathcal{L}_{\text{quant}} + \lambda \mathcal{L}_{\text{div}},   
\label{equ:total-loss}
\end{equation}
where $\mu \in [0, 1]$ and $\lambda \in [0, 1]$ are hyperparameters that control the contributions of the quantization loss and diversity loss, respectively. 
This combined loss function ensures that the model not only reconstructs the original POI embeddings accurately but also maintains semantic uniqueness and diversity in the generated SIDs.

\subsection{Generative POI Recommendation}
\label{fine-tune}

In the previous steps, we outline the process of converting POI data into SIDs. 
Building on these SIDs, we propose a generative POI recommendation framework based on LLMs. 
For a given user $u \in U$, we collect their historical check-in POI sequence $P_u = \{p_1, p_2, \dots, p_i\}$ and apply the semantic ID conversion method to transform this sequence into an SID sequence $S_u = \{s_1, s_2, \dots, s_i\}$. 
This transformation is formalized as:
\begin{equation}
    \mathcal{F}: P_u \rightarrow S_u,
\end{equation}
where $\mathcal{F}$ is the mapping function that converts POIs from their random numeric IDs to semantic IDs by our proposed method. 

After learning the SID of each POI, a straightforward approach is to integrate these SIDs into the LLM vocabulary, so that LLM can fulfill the Next POI recommendation task in a generative way. 
However, these SIDs are still OOV tokens for LLMs, making it necessary to align language tasks with recommendation tasks, thereby bridging the gap and ensuring effective knowledge transfer. 
To effectively bridge this gap, inspired by~\cite{Geng2022p5,Xue2022human,Zheng2024LCRec}, we initially construct a sequential POI prediction prompt that utilize the SID sequence of the user's historical check-in POIs in chronological order. The sequence of SIDs captures the user's evolving interests and behavioral characteristics over time.

Furthermore, considering that the temporal signals utilized in modeling the SID of a POI only reflect the frequent visitation patterns of the current POI (\ie collective temporal patterns exhibited by users), while the timestamp information in the user’s historical check-in records captures behavior-specific patterns of individual users (\eg habits, long-term preferences, and personalized temporal characteristics), we propose to chronologically concatenate the SIDs of historically visited POIs with their corresponding check-in timestamps. 
This integration facilitates a more comprehensive modeling of personalized user information and integrates collaborative semantics from both spatial and temporal dimensions.

Specifically, the prompt consists of two main components: (1) the instruction explicitly defines the task for the model, and (2) the input includes a user \textcolor{green}{uid} recent check-in history represented as an \textcolor{blue}{SID sequence}, the associated \textcolor{orange}{check-in times}, and the \textcolor{red}{current time} for prediction.
\begin{tcolorbox}[colback=yellow!20, colframe=white, coltitle=black, fonttitle=\bfseries, sharp corners]
 \textbf{Instruction:}  Here is a record of a user's POI accesses, your task is based on the history to predict the POI that the user is likely to access at the specified time.\\
 \textbf{Input:}  The user\_\textcolor{green}{<uid>} visited: \textcolor{blue}{<SID>} at \textcolor{orange}{[time]}, ..., visited \textcolor{blue}{<SID>} at \textcolor{orange}{[time]}. When \textcolor{red}{[time]} user\_\textcolor{green}{<uid>} is likely to visit:  
 \label{Prompt}
\end{tcolorbox}
This structured prompt enables the LLM to leverage both the semantic relationships encoded in the SIDs and the temporal patterns in the user's check-in history, facilitating accurate and context-aware next POI recommendations.
In contrast to existing works~\cite{li2024large}, which transforms the next POI recommendation into question-answering pairs in their prompt, we reformulate the problem as a fill-in-the-blank task. 
This task design enhances the model’s ability to identify and understand fine-grained temporal behaviors within specific time slots.

Since directly prompting an LLM often struggles to accurately capture the fine-grained relationships between historical check-in patterns and future POI recommendations~\cite{wang2023would}, to address this limitation, we fine-tune the LLM for the next POI recommendation task.
Specifically, we first create a supervised fine-tuning dataset, where we use the constructed prompt as the model input, and the SID of the POI actually visited by the user is used as the ground truth output. 
This approach enables the model to learn the mapping between historical check-in sequences and future POI visits, significantly enhancing the accuracy and relevance of the generated recommendations.

\section{Experimental Setting}
\label{settings}
 
\subsection{Datasets} 

We evaluate our approach on three real-world datasets: Foursquare-NYC~\cite{yang2014modeling}, Foursquare-TKY~\cite{yang2014modeling}, and Gowalla-CA~\cite{cho2011friendship}. 
Following \citet{yan2023spatio}, we preprocess each dataset by removing POIs with fewer than 10 interactions and users with fewer than 10 check-ins. 
The data is then sorted by time, with 80\% used for training, 10\% for validation, and 10\% for testing.
Users and POIs that do not appear in the training set are removed from the test set to ensure consistency between training and evaluation.
Subsequently, the check-in records are grouped by user and ordered chronologically. 
For each user, the last visited POI is held out as the ground truth during evaluation, while the preceding visits are used as the input sequence.
Note that in the training set, the user's visit records are also treated as historical sequences and concatenated with the test set to the specified length before being fed into the model.  
More detailed dataset statistics are provided in Table~\ref{tab: statistics}.

\begin{table}
  \caption{Statistics of three benchmark datasets.}
  \label{tab: statistics}
  \begin{tabular}{l>{\centering\arraybackslash}p{0.12\linewidth}>{\centering\arraybackslash}p{0.12\linewidth}>{\centering\arraybackslash}p{0.12\linewidth}}
    \toprule
    \textbf{Dataset}&\textbf{NYC}& \textbf{TKY} &\textbf{CA}\\
    \midrule
    Categories& 209& 190&301\\
    POIs& 5,135& 7,873&14,027\\
    Users& 1,083& 2,293&6,592\\
 Avg. Check-In & 136& 195&53
\\
    Fre. Check-In& 68& 100& 10\\
    \bottomrule
  \end{tabular}
\end{table}

\subsection{Implementation Detail} 

The quantization module consists of $3$ codebook layers, each containing $32$ or $64$ (32 for NYC and 64 for TKY and CA) codeword vectors with a dimensionality of $64$.
Following~\cite{zeghidour2021soundstream}, to prevent RQ-VAE from suffering the codebook space collapse we use k-means clustering-based initialization for each codebook layers. 
In Equation~\ref{equ:total-loss}, the weight of quantization loss $\mathcal{L}_{\text{quant}}$ and diversity loss $\mathcal{L}_{\text{div}}$ is set to $1.0$ and $0.25$ respectively.
For the generative recommendation module, we employ the LLaMA3-8B~\cite{hu2021lora} as the base model and fine-tune it using LoRA~\cite{dubey2024llama} with rank R 16 and dropout rate $0.1$. 
We employ a constant learning rate schedule with a learning rate of $1\text{e-}5$, combined with a warm-up phase of $20$ steps. 
The model is fine-tuned on $4$ Nvidia L40 GPUs, with a batch size of $2$ per GPU, gradient accumulation steps set to $8$, and a sequence length of $2,048$ tokens. 
Each input consists of the most recent $50$ check-in records of the user.

\subsection{Evaluation Metric \& Comparison Method}

Following previous work~\cite{feng2024rotan,li2024large}, we employ the accuracy metric for the top-1 recommendation at the specified time.
\begin{equation}
    \text{Acc@1} = \frac{1}{m} \sum_{i=1}^{m} \mathbb{I} (y_i = \hat{y}_i), 
\end{equation}
where $m$ is the number of test samples, $y_i$ and $\hat{y}_i$ represent the ground-truth label and the predict label of the $i$-th sample, respectively. 
$\mathbb{I}$ is the indicator function, which equals 1 when the condition is true and 0 otherwise.

We compare our proposed \model with several strong next POI recommendation baselines including the state-of-the-art LLM-based method.
(1) \textbf{Traditional methods}:
\texttt{PRME}~\cite{feng2015personalized}, a ranking-based metric embedding method; \texttt{LSTM}~\cite{graves2012long}, a sequential recommendation model based on RNNs; and \texttt{PLSPL}~\cite{wu2020personalized}, a personalized sequential recommendation model.
(2) \textbf{Transformer-based methods}:
\texttt{STAN}~\cite{luo2021stan}, a spatio-temporal attention network; and \texttt{GETNext}~\cite{yang2022getnext}, a Transformer-based next POI recommendation model.
(3) GCN-based method:
\texttt{STHGCN}~\cite{yan2023spatio}, a spatio-temporal hierarchical graph convolutional network.
(4) \textbf{Time-aware methods}:
\texttt{TPG}~\cite{luo2023timestamps}, a timestamp-guided model; and \texttt{ROTAN}~\cite{feng2024rotan}, a time-aware POI recommendation framework.
(5) \textbf{LLM-based method}:
\texttt{LLM4POI}~\cite{li2024large}, a state-of-the-art method based on LLMs which is the most advanced and closely related to our work. The key distinction is that \texttt{LLM4POI} relies on random numeric IDs, whereas our proposed \model leverages semantic codes.

\section{Experimental Result}

\subsection{Main Result}

\begin{table}
\centering
\caption{Comparison of different models on three datasets: NYC, TKY, and CA. We present the model inputs including the POI representation approach (POI Repre.) and whether visit timestamps are utilized.}
\begin{tabular}{lccccc}
\toprule
\multirow{2}{*}{\textbf{Model}} & \multicolumn{2}{c}{\textbf{Inputs}}& \multicolumn{3}{c}{\textbf{Acc@1}} \\ 
\cmidrule(lr){2-3} \cmidrule(lr){4-6}
& POI Repre. & Time & NYC& TKY& CA\\
\midrule
PRME
& RID& $\times$ & 0.1159
& 0.1052
& 0.0521
\\
LSTM
& RID& $\times$ & 0.1305& 0.1335
& 0.0665
\\
PLSPL
& RID& $\times$ & 0.1917& 0.1889
& 0.1072
\\
STAN
& RID& $\times$ & 0.2231& 0.1963
& 0.1104
\\
GETNext
& RID& $\times$ & 0.2435& 0.1829
& 0.1357
\\
STHGCN
& RID& $\times$ & 0.2734& 0.2950
& 0.1730
\\
TPG
& RID& $\checkmark$ & 0.2555
& 0.1420
& 0.1749
\\
ROTAN
& RID& $\checkmark$ & 0.3106
& 0.2458
& \underline{0.2199}\\

LLM4POI   & RID& $\checkmark$ & \underline{0.3372}& \underline{0.3035}& 0.2065
\\
\midrule
\textit{Llama3-8B}& RID& $\checkmark$ & 0.2643& 0.2123&0.1039\\
\textit{Llama3-8B}& SID& $\checkmark$ & 0.2885& 0.2282&0.1254\\
 \model& SID& $\checkmark$ & \textbf{0.3618}& \textbf{0.3062}& \textbf{0.2403}\\
 \bottomrule
\end{tabular}
\raggedright
\footnotesize {\textit{Llama3-8B} denotes the performance of Llama3-8B in a zero-shot setting, evaluated using both ID and SID, respectively.}
\label{tab:results}
\end{table}

In this section, we compare the performance of our proposed \model with various baselines on three benchmark datasets. 
The results shown in Table~\ref{tab:results} demonstrate that \model consistently outperforms all baselines across all datasets.
Specifically, \model achieves substantial improvements in top-1 accuracy compared to traditional model \texttt{ROTAN}, with relative gains of 16\%, 24\%, and 10\% on the NYC, TKY, and CA datasets, respectively 
These improvements underscore the efficacy of our proposed generative recommendation approach in the next POI recommendation task. 
This is attributed to the integration of the inherent commonsense knowledge and deep reasoning capabilities of LLMs into the POI recommendation task.
Furthermore, we also compare \model with the state-of-the-art LLM based method \texttt{LLM4POI}~\cite{li2024large}, which uses random numeric IDs to represent POIs. 
From the results we can find that \model achieves 7\%, 1\%, and 16\% of improvement compared with the SOTA method \texttt{LLM4POI} on the NYC, TKY, and CA datasets, respectively. 
This improvement demonstrate the effectiveness of our proposed semantic ID method which can capture more semantic information of POI than simply using the random numeric IDs.

\subsection{Ablation Study}
\label{ablation}

To verify the effectiveness of each module in our proposed \model, we conduct experiments on the following ablation models: 
(1) \textbf{w/o SID}: we directly use random number IDs instead of semantic ID (in Equation~\ref{equ:build-sid}), and keep the generative recommendation module.
(2) \textbf{w/o Time}: we remove the visit time (in \S~\ref{fine-tune}) from the users check-in records and use only the SID sequence as an input to LLM.
(3) \textbf{w/o $\mathcal{L}_{\text{div}}$}: we remove the diversity loss (in Equation~\ref{equ:diversity-loss}) during semantic ID construction.
(4) \textbf{w/o $\mathcal{L}_{\text{quant}}$}: we remove the quantization loss (in Equation~\ref{equ:quant-loss}) during semantic ID construction.
(5) \textbf{w/o $\mathcal{L}_{\text{recon}}$}: we remove the reconstruction loss (in Equation~\ref{equ:reconstruction-loss}) during semantic ID construction.

\begin{table}
\centering
\caption{Results of ablation models on NYC, TKY, and CA datasets with Acc@1 metric.}
\resizebox{1.0\columnwidth}{!}{
\begin{tabular}{lccc}
\toprule
\textbf{Ablations} &  \textbf{NYC} & \textbf{TKY} & \textbf{CA} \\
\midrule
\model & 0.3618& 0.3062 & 0.2403\\
w/o SID & 0.3356 ($\downarrow 7.24\%$)& 0.2812 ($\downarrow 8.18\%$)&0.2296($\downarrow 4.45\%$)\\
w/o Time & 0.2924 ($\downarrow19.18\%$)& 0.2568 ($\downarrow 16.13\%$)& 0.1971($\downarrow 17.8\%$)\\
w/o $\mathcal{L}_{\text{div}}$ & 0.3528 ($\downarrow 2.48\%$)& 0.2923 ($\downarrow 4.54\%$)&0.2313 ($\downarrow 3.72\%$)\\
w/o  $\mathcal{L}_{\text{quant}}$ & 0.3429 ($\downarrow 5.23\%$)& 0.2900 ($\downarrow 5.29\%$)& 0.2279($\downarrow 5.14\%$)\\
w/o  $\mathcal{L}_{\text{recon}}$ & ------& ------& ------\\
\bottomrule
\end{tabular}
}
\label{tab:ablation}
\end{table}

Table~\ref{tab:ablation} presents the Acc@1 performance of various ablation models across three datasets. 
In Table~\ref{tab:ablation}, "--" indicates that the ablation model collapsed during training and failed to converge. 
The results in Table~\ref{tab:ablation} demonstrate that the performance of these ablation models is lower than our proposed \model, thereby validating the effectiveness of all modules within \model. 
Notably, the model fails to converge when the $\mathcal{L}_{\text{recon}}$ loss function is removed. 
It is intuitive that the absence of constraints on the autoencoder output leads to the codebook being entirely incapable of modeling any semantic information of POIs.

\subsection{Analysis of Diversity Loss}

\begin{table}
\centering
\caption{Statistics of the constructed SID when using different weights $\lambda$ of diversity loss.}
\resizebox{1.0\columnwidth}{!}{
\begin{tabular}{lcccccc}
\toprule
\multirow{2}{*}{\textbf{$\lambda$}}& \multicolumn{3}{c}{\textbf{NYC}}& \multicolumn{3}{c}{\textbf{TKY}}\\ 
\cmidrule(lr){2-4} \cmidrule(lr){5-7}&Unique$\uparrow$& Collision$\downarrow$ &Acc@1& Unique$\uparrow$&Collision$\downarrow$  &Acc@1\\
\midrule
0.00&  2,916& 711&0.3528& 4,244&1,182 &0.2923\\
0.25& 3,259& 590  &\textbf{0.3618}& 4,465&1,147 &\textbf{0.3062}\\
 0.50& 3,292& 567  &0.3459& 4,481&1,126 &0.2898\\
0.75& 3,313& 554   &0.3368& 4,529&1,106 &0.2865\\
\bottomrule
\end{tabular}
}
\label{tab:diversity}
\end{table}

\begin{figure}
    \centering
    \subfloat[Category frequency distribution for different SID prefixes.]{
        \includegraphics[width=0.90\linewidth]{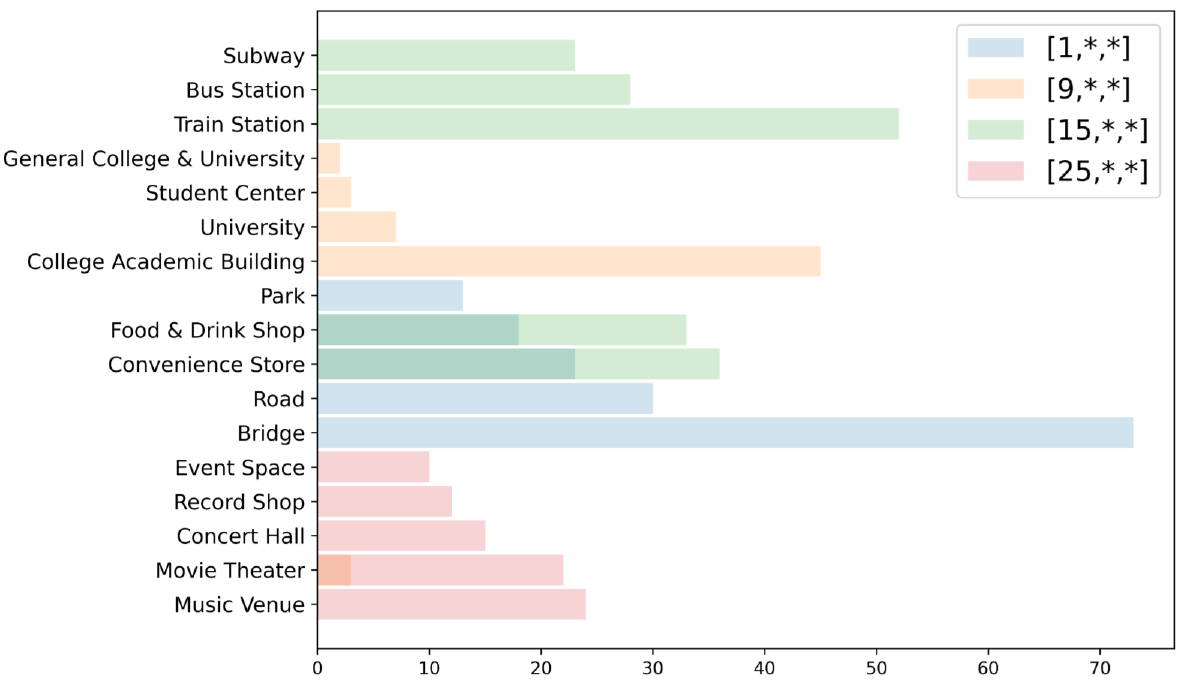}
        \label{fig:semantic-fig1}
    }
    \\
    \subfloat[The t-SNE visualization of SID.]{
        \includegraphics[width=0.90\linewidth]{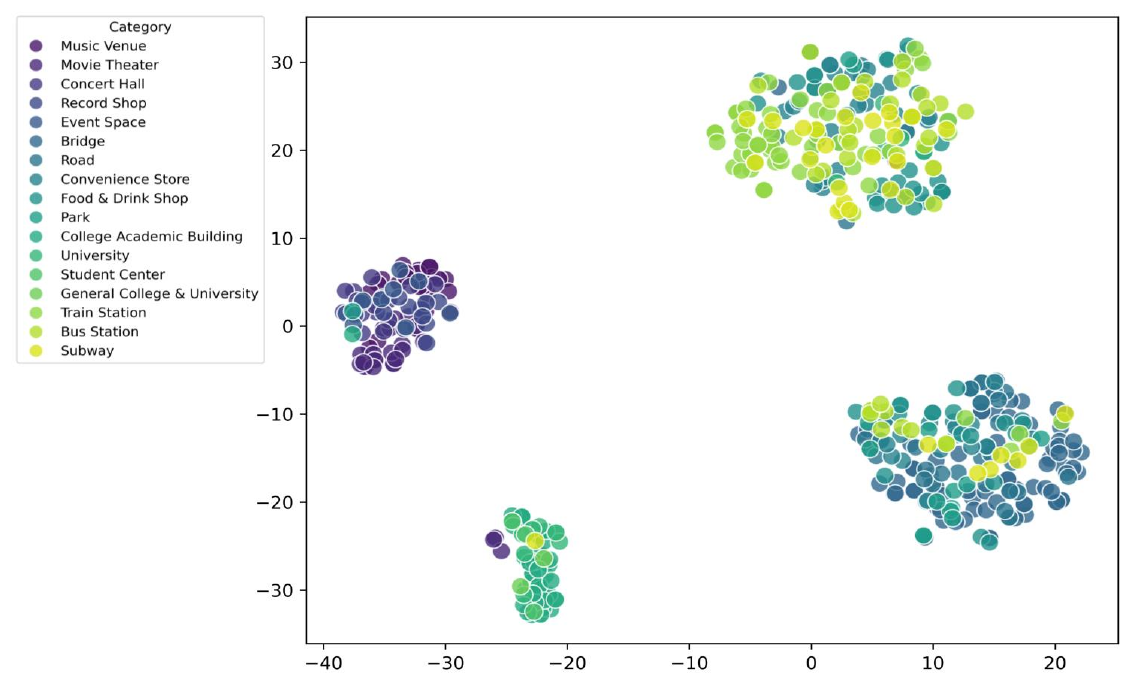}
        \label{fig:semantic-fig2}
    }
    \caption{SIDs semantic relevance analysis.}
    \label{fig:semantic}
\end{figure}

In this section, we investigate the effectiveness of our proposed diversity loss.
Table~\ref{tab:diversity} shows the model performance under different weight $\lambda$ of diversity loss when we combine this loss function to the total loss.
Recall that, we propose the diversity loss to prevent semantic collapse, where multiple POIs are mapped to the same SID. 
To quantitatively evaluate the impact of the diversity loss, we propose two metrics to assess the quality of the generated SIDs. 
First, we measure the number of unique SIDs generated by quantization. 
Second, we calculate the number of colliding SIDs (\ie multiple POIs map to the same SID ) generated by quantization. 

Table~\ref{tab:diversity} shows the model performance under different values of the hyper-parameter $\lambda$.
From Table~\ref{tab:diversity}, it can be observed that as the weight of the diversity loss increases, the collisions of semantic ID decreases. 
This observation demonstrates that the diversity loss effectively encourages the model to distribute POIs more uniformly across the semantic space. 
However, when the weight of the diversity loss becomes excessively large, the penalty forces the model to map the POI semantic vectors as uniformly as possible within the semantic space. 
This can result in similar POIs being assigned to different semantic clusters, thereby negatively impacting the performance of the next POI recommendation.

\subsection{Semantic Relevance Analysis of SID}

To intuitively validate whether the SIDs generated by our proposed method can effectively model the semantic information of POIs, we visualize the POIs in the TKY dataset. 
In Figure~\ref{fig:semantic-fig1}, we randomly select four SID prefixes with varying first codeword, and show the distribution of these SIDs across the corresponding POI categories.
From Figure~\ref{fig:semantic-fig1} we can find that different semantic spaces (\aka different SID prefixes) effectively capture information related to distinct semantic categories of POIs, while POIs of similar categories are quantized into the same semantic space (\aka the same SID prefix). 
For example, SIDs starting with <a\_15> represent transportation-related venues (\eg subways, bus stations, and train stations), while SIDs starting with <a\_25> correspond to entertainment venues (\eg music venues, record stores, and cinemas).

Furthermore, in Figure~\ref{fig:semantic-fig2}, we present the distribution of SIDs for POIs in the TKY dataset after t-SNE dimensionality reduction. 
The POIs are colored according to their categories as provided in the dataset. 
From Figure~\ref{fig:semantic-fig2}, it can be observed that POIs of different categories (\aka points of different colors) are well-separated, indicating significant distinguishability in their SIDs. 
In contrast, POIs of the same category (\aka points of the same color) are positioned closely. 
This experiment visually confirms that our proposed POI quantization method can map semantically similar POIs to adjacent SID spaces, which enables the LLM to better understand the semantic relationships between POIs and achieve improved generalization performance when encountering unseen POIs during inference.

\subsection{Performance on Out-of-domain Data}

\begin{table}
\centering
\caption{Comparison between \model and \texttt{LLM4POI} on out-of-domain dataset.}
\resizebox{1.0\columnwidth}{!}{
\begin{tabular}{lcccccc}
\toprule
\multirow{2}{*}{\textbf{Train Data}}&\multicolumn{3}{c}{\textbf{\model}} & \multicolumn{3}{c}{\textbf{LLM4POI}}\\ 
\cmidrule(lr){2-4} \cmidrule(lr){5-7}
&NYC& TKY& CA & NYC& TKY&CA \\
\midrule
NYC&0.3618& 0.2946& 0.2297& 0.3373& 0.2594&0.1885\\
TKY&0.3481& 0.3062
& 0.2342& 0.3463& 0.3035&0.1960\\
CA&0.3536& 0.2869& 0.2403& 0.3344& 0.2600&0.2065\\
\bottomrule
\end{tabular}
}
\label{tab:generalization}
\end{table}

In this paper, we propose the semantic ID construction module, independent of LLMs, to convert POI information into SIDs. 
In contrast, existing methods (\eg \texttt{LLM4POI}) employ random IDs, which forces the LLM to memorize the POI semantic information behind each random ID.
This approach leads to generalization issues when encountering new POIs, as the model fails to comprehend the semantic meaning of the random ID for new POI. 
Table~\ref{tab:generalization} shows the performance on out-of-domain datasets.
The performance degradation of \model on out-of-domain data is notably less pronounced compared to \texttt{LLM4POI} across all datasets, which further demonstrates that our proposed \model exhibits superior generalization capabilities over \texttt{LLM4POI}.

Especially in the model generalizes from short-sequence datasets (CA with 53 average check-ins) to long-sequence datasets (TKY with 195 average check-ins), our model \model outperforms \texttt{LLM4POI} by a significant margin.
Additionally, the generalization ability of \texttt{LLM4POI} heavily relies on the historical sequences of similar users. 
It feeds these sequences into the prompt to guide the model's predictions, requiring extra resources and computation, which is clearly unsuitable for out-of-domain data. 
In contrast, our model requires no additional prompts and achieves better performance than \texttt{LLM4POI} using only the current user's historical sequence. 
This further highlights the superior generalization capabilities of \model.

\subsection{Analysis of Model Efficiency}

\begin{table}
\centering
\caption{Efficiency comparison on the NYC dataset.}
\label{tab:efficiency}
\resizebox{1.0\columnwidth}{!}{
\begin{tabular}{cccc}
    \toprule
    \textbf{Models}&     \textbf{Training Time}& \textbf{Test Time}& \textbf{Total Tokens}\\
    \midrule
    LLM4POI &  5,108s& 230.36s& 5,061,008\\
    \model& 2,121s ($\downarrow58.48\%$)& 110.62s ($\downarrow51.98\%$)& 2,288,764($\downarrow54.78\%$)\\
    \bottomrule
\end{tabular}
}
\end{table}

In this paper, we propose the \model that uses SID, which includes POI semantic information, as the input to the LLM. 
Compared to methods using RID as input, our approach does not require the descriptive text of POIs to be fed into the LLM. 
This also leads to a significant reduction in the number of tokens in the LLM input of \model compared to methods that use RID.
Intuitively, this makes our method faster when predicting the next POIs using the generative model, especially when the historical check-in sequence is long. 
Table~\ref{tab:efficiency} presents a comparison of the efficiency between our SID-based \model and the RID-based method \texttt{LLM4POI}. 
We train both models for 8 epochs on the NYC dataset. 
As can be seen from Table~\ref{tab:efficiency}, since we compress all semantic information of the POI into the SID, the token usage in our \model is lower than that of \texttt{LLM4POI}, resulting in a reduction of 54.78\%. 
Additionally, our \model shows 51.98\% improvement in terms of test time. 
This result indicates that using the SID not only benefits the understanding of POI semantics and generates more accurate recommendations, but also improves the training and inference efficiency since we reduce the input tokens for the generative recommendation model.

\section{Conclusion}

In this paper, we proposed \fullmodel (\model), a novel LLM-based framework for next POI recommendation that uses the POI semantic ID to integrate more spatial-temporal information.
By designing a semantic ID construction module that captures POI semantics and a generative recommendation module that fine-tunes LLMs, \model significantly improves recommendation accuracy and generalizability. 
Our hierarchical POI quantization method ensures semantically similar POIs share similar IDs, while the diversity loss enhances codeword distinctiveness.
Extensive experiments on three benchmark datasets demonstrate that \model outperforms state-of-the-art methods, achieving accuracy improvements of up to 16\%. 
Semantic similarity analysis further confirms that our generated SIDs effectively capture POI semantic relationships. 

\begin{acks}
This work was supported by the National Key R\&D Program of China 2024YFE0111800, the National Natural Science Foundation of China (62432002, 62032001, 62406061, and T2293773), and the Natural Science Foundation of Shandong Province (ZR2023QF159).
\end{acks}

\bibliographystyle{ACM-Reference-Format}
\balance
\bibliography{reference}


\begin{thebibliography}{50}


\ifx \showCODEN    \undefined \def \showCODEN     #1{\unskip}     \fi
\ifx \showISBNx    \undefined \def \showISBNx     #1{\unskip}     \fi
\ifx \showISBNxiii \undefined \def \showISBNxiii  #1{\unskip}     \fi
\ifx \showISSN     \undefined \def \showISSN      #1{\unskip}     \fi
\ifx \showLCCN     \undefined \def \showLCCN      #1{\unskip}     \fi
\ifx \shownote     \undefined \def \shownote      #1{#1}          \fi
\ifx \showarticletitle \undefined \def \showarticletitle #1{#1}   \fi
\ifx \showURL      \undefined \def \showURL       {\relax}        \fi
\providecommand\bibfield[2]{#2}
\providecommand\bibinfo[2]{#2}
\providecommand\natexlab[1]{#1}
\providecommand\showeprint[2][]{arXiv:#2}

\bibitem[Bai et~al\mbox{.}(2023a)]%
        {bai2023qwen}
\bibfield{author}{\bibinfo{person}{Jinze Bai}, \bibinfo{person}{Shuai Bai}, \bibinfo{person}{Yunfei Chu}, \bibinfo{person}{Zeyu Cui}, \bibinfo{person}{Kai Dang}, \bibinfo{person}{Xiaodong Deng}, \bibinfo{person}{Yang Fan}, \bibinfo{person}{Wenbin Ge}, \bibinfo{person}{Yu Han}, \bibinfo{person}{Fei Huang}, {et~al\mbox{.}}} \bibinfo{year}{2023}\natexlab{a}.
\newblock \showarticletitle{Qwen technical report}.
\newblock \bibinfo{journal}{\emph{arXiv preprint arXiv:2309.16609}} (\bibinfo{year}{2023}).
\newblock


\bibitem[Bai et~al\mbox{.}(2023b)]%
        {Bai2023QwenVL}
\bibfield{author}{\bibinfo{person}{Jinze Bai}, \bibinfo{person}{Shuai Bai}, \bibinfo{person}{Shusheng Yang}, \bibinfo{person}{Shijie Wang}, \bibinfo{person}{Sinan Tan}, \bibinfo{person}{Peng Wang}, \bibinfo{person}{Junyang Lin}, \bibinfo{person}{Chang Zhou}, {and} \bibinfo{person}{Jingren Zhou}.} \bibinfo{year}{2023}\natexlab{b}.
\newblock \showarticletitle{Qwen-VL: A Versatile Vision-Language Model for Understanding, Localization, Text Reading, and Beyond}.
\newblock \bibinfo{journal}{\emph{arXiv preprint arXiv:2308.12966}} (\bibinfo{year}{2023}).
\newblock


\bibitem[Brown et~al\mbox{.}(2020)]%
        {brown2020language}
\bibfield{author}{\bibinfo{person}{Tom Brown}, \bibinfo{person}{Benjamin Mann}, \bibinfo{person}{Nick Ryder}, \bibinfo{person}{Melanie Subbiah}, \bibinfo{person}{Jared~D Kaplan}, \bibinfo{person}{Prafulla Dhariwal}, \bibinfo{person}{Arvind Neelakantan}, \bibinfo{person}{Pranav Shyam}, \bibinfo{person}{Girish Sastry}, \bibinfo{person}{Amanda Askell}, {et~al\mbox{.}}} \bibinfo{year}{2020}\natexlab{}.
\newblock \showarticletitle{Language models are few-shot learners}.
\newblock \bibinfo{journal}{\emph{Advances in neural information processing systems}}  \bibinfo{volume}{33} (\bibinfo{year}{2020}), \bibinfo{pages}{1877--1901}.
\newblock


\bibitem[Cheng et~al\mbox{.}(2013)]%
        {cheng2013you}
\bibfield{author}{\bibinfo{person}{Chen Cheng}, \bibinfo{person}{Haiqin Yang}, \bibinfo{person}{Michael~R Lyu}, {and} \bibinfo{person}{Irwin King}.} \bibinfo{year}{2013}\natexlab{}.
\newblock \showarticletitle{Where you like to go next: Successive point-of-interest recommendation}. In \bibinfo{booktitle}{\emph{Twenty-Third international joint conference on Artificial Intelligence}}.
\newblock


\bibitem[Cho et~al\mbox{.}(2011)]%
        {cho2011friendship}
\bibfield{author}{\bibinfo{person}{Eunjoon Cho}, \bibinfo{person}{Seth~A Myers}, {and} \bibinfo{person}{Jure Leskovec}.} \bibinfo{year}{2011}\natexlab{}.
\newblock \showarticletitle{Friendship and mobility: user movement in location-based social networks}. In \bibinfo{booktitle}{\emph{Proceedings of the 17th ACM SIGKDD international conference on Knowledge discovery and data mining}}. \bibinfo{pages}{1082--1090}.
\newblock


\bibitem[Duan et~al\mbox{.}(2023)]%
        {duan2023clsprec}
\bibfield{author}{\bibinfo{person}{Chenghua Duan}, \bibinfo{person}{Wei Fan}, \bibinfo{person}{Wei Zhou}, \bibinfo{person}{Hu Liu}, {and} \bibinfo{person}{Junhao Wen}.} \bibinfo{year}{2023}\natexlab{}.
\newblock \showarticletitle{Clsprec: Contrastive learning of long and short-term preferences for next poi recommendation}. In \bibinfo{booktitle}{\emph{Proceedings of the 32nd acm international conference on information and knowledge management}}. \bibinfo{pages}{473--482}.
\newblock


\bibitem[Dubey et~al\mbox{.}(2024)]%
        {dubey2024llama}
\bibfield{author}{\bibinfo{person}{Abhimanyu Dubey}, \bibinfo{person}{Abhinav Jauhri}, \bibinfo{person}{Abhinav Pandey}, \bibinfo{person}{Abhishek Kadian}, \bibinfo{person}{Ahmad Al-Dahle}, \bibinfo{person}{Aiesha Letman}, \bibinfo{person}{Akhil Mathur}, \bibinfo{person}{Alan Schelten}, \bibinfo{person}{Amy Yang}, \bibinfo{person}{Angela Fan}, {et~al\mbox{.}}} \bibinfo{year}{2024}\natexlab{}.
\newblock \showarticletitle{The llama 3 herd of models}.
\newblock \bibinfo{journal}{\emph{arXiv preprint arXiv:2407.21783}} (\bibinfo{year}{2024}).
\newblock


\bibitem[Feng et~al\mbox{.}(2015)]%
        {feng2015personalized}
\bibfield{author}{\bibinfo{person}{Shanshan Feng}, \bibinfo{person}{Xutao Li}, \bibinfo{person}{Yifeng Zeng}, \bibinfo{person}{Gao Cong}, \bibinfo{person}{Yeow~Meng Chee}, {and} \bibinfo{person}{Quan Yuan}.} \bibinfo{year}{2015}\natexlab{}.
\newblock \showarticletitle{Personalized ranking metric embedding for next new poi recommendation}.
\newblock  (\bibinfo{year}{2015}).
\newblock


\bibitem[Feng et~al\mbox{.}(2024a)]%
        {feng2024move}
\bibfield{author}{\bibinfo{person}{Shanshan Feng}, \bibinfo{person}{Haoming Lyu}, \bibinfo{person}{Fan Li}, \bibinfo{person}{Zhu Sun}, {and} \bibinfo{person}{Caishun Chen}.} \bibinfo{year}{2024}\natexlab{a}.
\newblock \showarticletitle{Where to move next: Zero-shot generalization of llms for next poi recommendation}. In \bibinfo{booktitle}{\emph{2024 IEEE Conference on Artificial Intelligence (CAI)}}. IEEE, \bibinfo{pages}{1530--1535}.
\newblock


\bibitem[Feng et~al\mbox{.}(2024b)]%
        {feng2024rotan}
\bibfield{author}{\bibinfo{person}{Shanshan Feng}, \bibinfo{person}{Feiyu Meng}, \bibinfo{person}{Lisi Chen}, \bibinfo{person}{Shuo Shang}, {and} \bibinfo{person}{Yew~Soon Ong}.} \bibinfo{year}{2024}\natexlab{b}.
\newblock \showarticletitle{Rotan: A rotation-based temporal attention network for time-specific next poi recommendation}. In \bibinfo{booktitle}{\emph{Proceedings of the 30th ACM SIGKDD Conference on Knowledge Discovery and Data Mining}}. \bibinfo{pages}{759--770}.
\newblock


\bibitem[Geng et~al\mbox{.}(2022)]%
        {Geng2022p5}
\bibfield{author}{\bibinfo{person}{Shijie Geng}, \bibinfo{person}{Shuchang Liu}, \bibinfo{person}{Zuohui Fu}, \bibinfo{person}{Yingqiang Ge}, {and} \bibinfo{person}{Yongfeng Zhang}.} \bibinfo{year}{2022}\natexlab{}.
\newblock \showarticletitle{Recommendation as Language Processing (RLP): A Unified Pretrain, Personalized Prompt \& Predict Paradigm (P5)}. In \bibinfo{booktitle}{\emph{Proceedings of the 16th ACM Conference on Recommender Systems}}. \bibinfo{pages}{299–315}.
\newblock


\bibitem[Graves and Graves(2012)]%
        {graves2012long}
\bibfield{author}{\bibinfo{person}{Alex Graves} {and} \bibinfo{person}{Alex Graves}.} \bibinfo{year}{2012}\natexlab{}.
\newblock \showarticletitle{Long short-term memory}.
\newblock \bibinfo{journal}{\emph{Supervised sequence labelling with recurrent neural networks}} (\bibinfo{year}{2012}), \bibinfo{pages}{37--45}.
\newblock


\bibitem[Han et~al\mbox{.}(2024)]%
        {Han2024OneLLM}
\bibfield{author}{\bibinfo{person}{Jiaming Han}, \bibinfo{person}{Kaixiong Gong}, \bibinfo{person}{Yiyuan Zhang}, \bibinfo{person}{Jiaqi Wang}, \bibinfo{person}{Kaipeng Zhang}, \bibinfo{person}{Dahua Lin}, \bibinfo{person}{Yu Qiao}, \bibinfo{person}{Peng Gao}, {and} \bibinfo{person}{Xiangyu Yue}.} \bibinfo{year}{2024}\natexlab{}.
\newblock \showarticletitle{OneLLM: One Framework to Align All Modalities with Language}. In \bibinfo{booktitle}{\emph{Proceedings of the IEEE/CVF Conference on Computer Vision and Pattern Recognition (CVPR)}}. \bibinfo{pages}{26584--26595}.
\newblock


\bibitem[He et~al\mbox{.}(2016)]%
        {he2016inferring}
\bibfield{author}{\bibinfo{person}{Jing He}, \bibinfo{person}{Xin Li}, \bibinfo{person}{Lejian Liao}, \bibinfo{person}{Dandan Song}, {and} \bibinfo{person}{William Cheung}.} \bibinfo{year}{2016}\natexlab{}.
\newblock \showarticletitle{Inferring a personalized next point-of-interest recommendation model with latent behavior patterns}. In \bibinfo{booktitle}{\emph{Proceedings of the AAAI conference on artificial intelligence}}, Vol.~\bibinfo{volume}{30}.
\newblock


\bibitem[Hu et~al\mbox{.}(2021)]%
        {hu2021lora}
\bibfield{author}{\bibinfo{person}{Edward~J Hu}, \bibinfo{person}{Yelong Shen}, \bibinfo{person}{Phillip Wallis}, \bibinfo{person}{Zeyuan Allen-Zhu}, \bibinfo{person}{Yuanzhi Li}, \bibinfo{person}{Shean Wang}, \bibinfo{person}{Lu Wang}, {and} \bibinfo{person}{Weizhu Chen}.} \bibinfo{year}{2021}\natexlab{}.
\newblock \showarticletitle{Lora: Low-rank adaptation of large language models}.
\newblock \bibinfo{journal}{\emph{arXiv preprint arXiv:2106.09685}} (\bibinfo{year}{2021}).
\newblock


\bibitem[Islam et~al\mbox{.}(2022)]%
        {islam2022survey}
\bibfield{author}{\bibinfo{person}{Md~Ashraful Islam}, \bibinfo{person}{Mir~Mahathir Mohammad}, \bibinfo{person}{Sarkar Snigdha~Sarathi Das}, {and} \bibinfo{person}{Mohammed~Eunus Ali}.} \bibinfo{year}{2022}\natexlab{}.
\newblock \showarticletitle{A survey on deep learning based Point-of-Interest (POI) recommendations}.
\newblock \bibinfo{journal}{\emph{Neurocomputing}}  \bibinfo{volume}{472} (\bibinfo{year}{2022}), \bibinfo{pages}{306--325}.
\newblock


\bibitem[Kong and Wu(2018)]%
        {kong2018hst}
\bibfield{author}{\bibinfo{person}{Dejiang Kong} {and} \bibinfo{person}{Fei Wu}.} \bibinfo{year}{2018}\natexlab{}.
\newblock \showarticletitle{HST-LSTM: A hierarchical spatial-temporal long-short term memory network for location prediction.}. In \bibinfo{booktitle}{\emph{Ijcai}}, Vol.~\bibinfo{volume}{18}. \bibinfo{pages}{2341--2347}.
\newblock


\bibitem[Li et~al\mbox{.}(2024)]%
        {li2024large}
\bibfield{author}{\bibinfo{person}{Peibo Li}, \bibinfo{person}{Maarten de Rijke}, \bibinfo{person}{Hao Xue}, \bibinfo{person}{Shuang Ao}, \bibinfo{person}{Yang Song}, {and} \bibinfo{person}{Flora~D Salim}.} \bibinfo{year}{2024}\natexlab{}.
\newblock \showarticletitle{Large language models for next point-of-interest recommendation}. In \bibinfo{booktitle}{\emph{Proceedings of the 47th International ACM SIGIR Conference on Research and Development in Information Retrieval}}. \bibinfo{pages}{1463--1472}.
\newblock


\bibitem[Lian et~al\mbox{.}(2020)]%
        {lian2020geography}
\bibfield{author}{\bibinfo{person}{Defu Lian}, \bibinfo{person}{Yongji Wu}, \bibinfo{person}{Yong Ge}, \bibinfo{person}{Xing Xie}, {and} \bibinfo{person}{Enhong Chen}.} \bibinfo{year}{2020}\natexlab{}.
\newblock \showarticletitle{Geography-aware sequential location recommendation}. In \bibinfo{booktitle}{\emph{Proceedings of the 26th ACM SIGKDD international conference on knowledge discovery \& data mining}}. \bibinfo{pages}{2009--2019}.
\newblock


\bibitem[Lim et~al\mbox{.}(2022)]%
        {lim2022hierarchical}
\bibfield{author}{\bibinfo{person}{Nicholas Lim}, \bibinfo{person}{Bryan Hooi}, \bibinfo{person}{See-Kiong Ng}, \bibinfo{person}{Yong~Liang Goh}, \bibinfo{person}{Renrong Weng}, {and} \bibinfo{person}{Rui Tan}.} \bibinfo{year}{2022}\natexlab{}.
\newblock \showarticletitle{Hierarchical multi-task graph recurrent network for next poi recommendation}. In \bibinfo{booktitle}{\emph{Proceedings of the 45th international ACM SIGIR conference on Research and development in Information Retrieval}}. \bibinfo{pages}{1133--1143}.
\newblock


\bibitem[Lim et~al\mbox{.}(2020)]%
        {lim2020stp}
\bibfield{author}{\bibinfo{person}{Nicholas Lim}, \bibinfo{person}{Bryan Hooi}, \bibinfo{person}{See-Kiong Ng}, \bibinfo{person}{Xueou Wang}, \bibinfo{person}{Yong~Liang Goh}, \bibinfo{person}{Renrong Weng}, {and} \bibinfo{person}{Jagannadan Varadarajan}.} \bibinfo{year}{2020}\natexlab{}.
\newblock \showarticletitle{STP-UDGAT: Spatial-temporal-preference user dimensional graph attention network for next POI recommendation}. In \bibinfo{booktitle}{\emph{Proceedings of the 29th ACM International conference on information \& knowledge management}}. \bibinfo{pages}{845--854}.
\newblock


\bibitem[Liu et~al\mbox{.}(2016)]%
        {liu2016predicting}
\bibfield{author}{\bibinfo{person}{Qiang Liu}, \bibinfo{person}{Shu Wu}, \bibinfo{person}{Liang Wang}, {and} \bibinfo{person}{Tieniu Tan}.} \bibinfo{year}{2016}\natexlab{}.
\newblock \showarticletitle{Predicting the next location: A recurrent model with spatial and temporal contexts}. In \bibinfo{booktitle}{\emph{Proceedings of the AAAI conference on artificial intelligence}}, Vol.~\bibinfo{volume}{30}.
\newblock


\bibitem[Liu et~al\mbox{.}(2024)]%
        {liu2024nextlocllm}
\bibfield{author}{\bibinfo{person}{Shuai Liu}, \bibinfo{person}{Ning Cao}, \bibinfo{person}{Yile Chen}, \bibinfo{person}{Yue Jiang}, {and} \bibinfo{person}{Gao Cong}.} \bibinfo{year}{2024}\natexlab{}.
\newblock \showarticletitle{nextlocllm: next location prediction using LLMs}.
\newblock \bibinfo{journal}{\emph{arXiv preprint arXiv:2410.09129}} (\bibinfo{year}{2024}).
\newblock


\bibitem[Liu et~al\mbox{.}(2017)]%
        {liu2017experimental}
\bibfield{author}{\bibinfo{person}{Yiding Liu}, \bibinfo{person}{Tuan-Anh~Nguyen Pham}, \bibinfo{person}{Gao Cong}, {and} \bibinfo{person}{Quan Yuan}.} \bibinfo{year}{2017}\natexlab{}.
\newblock \showarticletitle{An experimental evaluation of point-of-interest recommendation in location-based social networks}.
\newblock  (\bibinfo{year}{2017}).
\newblock


\bibitem[Luo et~al\mbox{.}(2023)]%
        {luo2023timestamps}
\bibfield{author}{\bibinfo{person}{Yan Luo}, \bibinfo{person}{Haoyi Duan}, \bibinfo{person}{Ye Liu}, {and} \bibinfo{person}{Fu-Lai Chung}.} \bibinfo{year}{2023}\natexlab{}.
\newblock \showarticletitle{Timestamps as Prompts for Geography-Aware Location Recommendation}. In \bibinfo{booktitle}{\emph{Proceedings of the 32nd ACM International Conference on Information and Knowledge Management}}. \bibinfo{pages}{1697--1706}.
\newblock


\bibitem[Luo et~al\mbox{.}(2021)]%
        {luo2021stan}
\bibfield{author}{\bibinfo{person}{Yingtao Luo}, \bibinfo{person}{Qiang Liu}, {and} \bibinfo{person}{Zhaocheng Liu}.} \bibinfo{year}{2021}\natexlab{}.
\newblock \showarticletitle{Stan: Spatio-temporal attention network for next location recommendation}. In \bibinfo{booktitle}{\emph{Proceedings of the web conference 2021}}. \bibinfo{pages}{2177--2185}.
\newblock


\bibitem[Rajput et~al\mbox{.}(2023)]%
        {rajput2023recommender}
\bibfield{author}{\bibinfo{person}{Shashank Rajput}, \bibinfo{person}{Nikhil Mehta}, \bibinfo{person}{Anima Singh}, \bibinfo{person}{Raghunandan Hulikal~Keshavan}, \bibinfo{person}{Trung Vu}, \bibinfo{person}{Lukasz Heldt}, \bibinfo{person}{Lichan Hong}, \bibinfo{person}{Yi Tay}, \bibinfo{person}{Vinh Tran}, \bibinfo{person}{Jonah Samost}, {et~al\mbox{.}}} \bibinfo{year}{2023}\natexlab{}.
\newblock \showarticletitle{Recommender systems with generative retrieval}.
\newblock \bibinfo{journal}{\emph{Advances in Neural Information Processing Systems}}  \bibinfo{volume}{36} (\bibinfo{year}{2023}), \bibinfo{pages}{10299--10315}.
\newblock


\bibitem[S{\'a}nchez and Bellog{\'\i}n(2022)]%
        {sanchez2022point}
\bibfield{author}{\bibinfo{person}{Pablo S{\'a}nchez} {and} \bibinfo{person}{Alejandro Bellog{\'\i}n}.} \bibinfo{year}{2022}\natexlab{}.
\newblock \showarticletitle{Point-of-interest recommender systems based on location-based social networks: a survey from an experimental perspective}.
\newblock \bibinfo{journal}{\emph{ACM Computing Surveys (CSUR)}} \bibinfo{volume}{54}, \bibinfo{number}{11s} (\bibinfo{year}{2022}), \bibinfo{pages}{1--37}.
\newblock


\bibitem[Sun et~al\mbox{.}(2020)]%
        {sun2020go}
\bibfield{author}{\bibinfo{person}{Ke Sun}, \bibinfo{person}{Tieyun Qian}, \bibinfo{person}{Tong Chen}, \bibinfo{person}{Yile Liang}, \bibinfo{person}{Quoc Viet~Hung Nguyen}, {and} \bibinfo{person}{Hongzhi Yin}.} \bibinfo{year}{2020}\natexlab{}.
\newblock \showarticletitle{Where to go next: Modeling long-and short-term user preferences for point-of-interest recommendation}. In \bibinfo{booktitle}{\emph{Proceedings of the AAAI conference on artificial intelligence}}, Vol.~\bibinfo{volume}{34}. \bibinfo{pages}{214--221}.
\newblock


\bibitem[Sun et~al\mbox{.}(2024)]%
        {sun2024learning}
\bibfield{author}{\bibinfo{person}{Weiwei Sun}, \bibinfo{person}{Lingyong Yan}, \bibinfo{person}{Zheng Chen}, \bibinfo{person}{Shuaiqiang Wang}, \bibinfo{person}{Haichao Zhu}, \bibinfo{person}{Pengjie Ren}, \bibinfo{person}{Zhumin Chen}, \bibinfo{person}{Dawei Yin}, \bibinfo{person}{Maarten Rijke}, {and} \bibinfo{person}{Zhaochun Ren}.} \bibinfo{year}{2024}\natexlab{}.
\newblock \showarticletitle{Learning to tokenize for generative retrieval}.
\newblock \bibinfo{journal}{\emph{Advances in Neural Information Processing Systems}}  \bibinfo{volume}{36} (\bibinfo{year}{2024}).
\newblock


\bibitem[Van Den~Oord et~al\mbox{.}(2017)]%
        {van2017neural}
\bibfield{author}{\bibinfo{person}{Aaron Van Den~Oord}, \bibinfo{person}{Oriol Vinyals}, {et~al\mbox{.}}} \bibinfo{year}{2017}\natexlab{}.
\newblock \showarticletitle{Neural discrete representation learning}.
\newblock \bibinfo{journal}{\emph{Advances in neural information processing systems}}  \bibinfo{volume}{30} (\bibinfo{year}{2017}).
\newblock


\bibitem[Wang et~al\mbox{.}(2022b)]%
        {wang2022spatial}
\bibfield{author}{\bibinfo{person}{En Wang}, \bibinfo{person}{Yiheng Jiang}, \bibinfo{person}{Yuanbo Xu}, \bibinfo{person}{Liang Wang}, {and} \bibinfo{person}{Yongjian Yang}.} \bibinfo{year}{2022}\natexlab{b}.
\newblock \showarticletitle{Spatial-temporal interval aware sequential POI recommendation}. In \bibinfo{booktitle}{\emph{2022 IEEE 38th international conference on data engineering (ICDE)}}. IEEE, \bibinfo{pages}{2086--2098}.
\newblock


\bibitem[Wang et~al\mbox{.}(2024b)]%
        {wang2024large}
\bibfield{author}{\bibinfo{person}{Jiawei Wang}, \bibinfo{person}{Renhe Jiang}, \bibinfo{person}{Chuang Yang}, \bibinfo{person}{Zengqing Wu}, \bibinfo{person}{Makoto Onizuka}, \bibinfo{person}{Ryosuke Shibasaki}, \bibinfo{person}{Noboru Koshizuka}, {and} \bibinfo{person}{Chuan Xiao}.} \bibinfo{year}{2024}\natexlab{b}.
\newblock \showarticletitle{Large language models as urban residents: An llm agent framework for personal mobility generation}.
\newblock \bibinfo{journal}{\emph{arXiv preprint arXiv:2402.14744}} (\bibinfo{year}{2024}).
\newblock


\bibitem[Wang et~al\mbox{.}(2024a)]%
        {wang2024learnable}
\bibfield{author}{\bibinfo{person}{Wenjie Wang}, \bibinfo{person}{Honghui Bao}, \bibinfo{person}{Xinyu Lin}, \bibinfo{person}{Jizhi Zhang}, \bibinfo{person}{Yongqi Li}, \bibinfo{person}{Fuli Feng}, \bibinfo{person}{See-Kiong Ng}, {and} \bibinfo{person}{Tat-Seng Chua}.} \bibinfo{year}{2024}\natexlab{a}.
\newblock \showarticletitle{Learnable item tokenization for generative recommendation}. In \bibinfo{booktitle}{\emph{Proceedings of the 33rd ACM International Conference on Information and Knowledge Management}}. \bibinfo{pages}{2400--2409}.
\newblock


\bibitem[Wang et~al\mbox{.}(2023a)]%
        {wang2023would}
\bibfield{author}{\bibinfo{person}{Xinglei Wang}, \bibinfo{person}{Meng Fang}, \bibinfo{person}{Zichao Zeng}, {and} \bibinfo{person}{Tao Cheng}.} \bibinfo{year}{2023}\natexlab{a}.
\newblock \showarticletitle{Where would i go next? large language models as human mobility predictors}.
\newblock \bibinfo{journal}{\emph{arXiv preprint arXiv:2308.15197}} (\bibinfo{year}{2023}).
\newblock


\bibitem[Wang et~al\mbox{.}(2022a)]%
        {wang2022neural}
\bibfield{author}{\bibinfo{person}{Yujing Wang}, \bibinfo{person}{Yingyan Hou}, \bibinfo{person}{Haonan Wang}, \bibinfo{person}{Ziming Miao}, \bibinfo{person}{Shibin Wu}, \bibinfo{person}{Qi Chen}, \bibinfo{person}{Yuqing Xia}, \bibinfo{person}{Chengmin Chi}, \bibinfo{person}{Guoshuai Zhao}, \bibinfo{person}{Zheng Liu}, {et~al\mbox{.}}} \bibinfo{year}{2022}\natexlab{a}.
\newblock \showarticletitle{A neural corpus indexer for document retrieval}.
\newblock \bibinfo{journal}{\emph{Advances in Neural Information Processing Systems}}  \bibinfo{volume}{35} (\bibinfo{year}{2022}), \bibinfo{pages}{25600--25614}.
\newblock


\bibitem[Wang et~al\mbox{.}(2024c)]%
        {wang2024enhanced}
\bibfield{author}{\bibinfo{person}{Yidan Wang}, \bibinfo{person}{Zhaochun Ren}, \bibinfo{person}{Weiwei Sun}, \bibinfo{person}{Jiyuan Yang}, \bibinfo{person}{Zhixiang Liang}, \bibinfo{person}{Xin Chen}, \bibinfo{person}{Ruobing Xie}, \bibinfo{person}{Su Yan}, \bibinfo{person}{Xu Zhang}, \bibinfo{person}{Pengjie Ren}, {et~al\mbox{.}}} \bibinfo{year}{2024}\natexlab{c}.
\newblock \showarticletitle{Enhanced generative recommendation via content and collaboration integration}.
\newblock \bibinfo{journal}{\emph{arXiv preprint arXiv:2403.18480}} (\bibinfo{year}{2024}).
\newblock


\bibitem[Wang et~al\mbox{.}(2022c)]%
        {wang2022learning}
\bibfield{author}{\bibinfo{person}{Zhaobo Wang}, \bibinfo{person}{Yanmin Zhu}, \bibinfo{person}{Haobing Liu}, {and} \bibinfo{person}{Chunyang Wang}.} \bibinfo{year}{2022}\natexlab{c}.
\newblock \showarticletitle{Learning graph-based disentangled representations for next POI recommendation}. In \bibinfo{booktitle}{\emph{Proceedings of the 45th international ACM SIGIR conference on research and development in information retrieval}}. \bibinfo{pages}{1154--1163}.
\newblock


\bibitem[Wang et~al\mbox{.}(2023b)]%
        {wang2023adaptive}
\bibfield{author}{\bibinfo{person}{Zhaobo Wang}, \bibinfo{person}{Yanmin Zhu}, \bibinfo{person}{Chunyang Wang}, \bibinfo{person}{Wenze Ma}, \bibinfo{person}{Bo Li}, {and} \bibinfo{person}{Jiadi Yu}.} \bibinfo{year}{2023}\natexlab{b}.
\newblock \showarticletitle{Adaptive Graph Representation Learning for Next POI Recommendation}. In \bibinfo{booktitle}{\emph{Proceedings of the 46th International ACM SIGIR Conference on Research and Development in Information Retrieval}}. \bibinfo{pages}{393--402}.
\newblock


\bibitem[Wongso et~al\mbox{.}(2024)]%
        {wongso2024genup}
\bibfield{author}{\bibinfo{person}{Wilson Wongso}, \bibinfo{person}{Hao Xue}, {and} \bibinfo{person}{Flora~D Salim}.} \bibinfo{year}{2024}\natexlab{}.
\newblock \showarticletitle{GenUP: Generative User Profilers as In-Context Learners for Next POI Recommender Systems}.
\newblock \bibinfo{journal}{\emph{arXiv preprint arXiv:2410.20643}} (\bibinfo{year}{2024}).
\newblock


\bibitem[Wu et~al\mbox{.}(2020)]%
        {wu2020personalized}
\bibfield{author}{\bibinfo{person}{Yuxia Wu}, \bibinfo{person}{Ke Li}, \bibinfo{person}{Guoshuai Zhao}, {and} \bibinfo{person}{Xueming Qian}.} \bibinfo{year}{2020}\natexlab{}.
\newblock \showarticletitle{Personalized long-and short-term preference learning for next POI recommendation}.
\newblock \bibinfo{journal}{\emph{IEEE Transactions on Knowledge and Data Engineering}} \bibinfo{volume}{34}, \bibinfo{number}{4} (\bibinfo{year}{2020}), \bibinfo{pages}{1944--1957}.
\newblock


\bibitem[Xue et~al\mbox{.}(2022)]%
        {Xue2022human}
\bibfield{author}{\bibinfo{person}{Hao Xue}, \bibinfo{person}{Bhanu~Prakash Voutharoja}, {and} \bibinfo{person}{Flora~D. Salim}.} \bibinfo{year}{2022}\natexlab{}.
\newblock \showarticletitle{Leveraging language foundation models for human mobility forecasting}. In \bibinfo{booktitle}{\emph{Proceedings of the 30th International Conference on Advances in Geographic Information Systems}}. \bibinfo{pages}{9}.
\newblock


\bibitem[Yan et~al\mbox{.}(2023)]%
        {yan2023spatio}
\bibfield{author}{\bibinfo{person}{Xiaodong Yan}, \bibinfo{person}{Tengwei Song}, \bibinfo{person}{Yifeng Jiao}, \bibinfo{person}{Jianshan He}, \bibinfo{person}{Jiaotuan Wang}, \bibinfo{person}{Ruopeng Li}, {and} \bibinfo{person}{Wei Chu}.} \bibinfo{year}{2023}\natexlab{}.
\newblock \showarticletitle{Spatio-temporal hypergraph learning for next POI recommendation}. In \bibinfo{booktitle}{\emph{Proceedings of the 46th international ACM SIGIR conference on research and development in information retrieval}}. \bibinfo{pages}{403--412}.
\newblock


\bibitem[Yang et~al\mbox{.}(2014)]%
        {yang2014modeling}
\bibfield{author}{\bibinfo{person}{Dingqi Yang}, \bibinfo{person}{Daqing Zhang}, \bibinfo{person}{Vincent~W Zheng}, {and} \bibinfo{person}{Zhiyong Yu}.} \bibinfo{year}{2014}\natexlab{}.
\newblock \showarticletitle{Modeling user activity preference by leveraging user spatial temporal characteristics in LBSNs}.
\newblock \bibinfo{journal}{\emph{IEEE Transactions on Systems, Man, and Cybernetics: Systems}} \bibinfo{volume}{45}, \bibinfo{number}{1} (\bibinfo{year}{2014}), \bibinfo{pages}{129--142}.
\newblock


\bibitem[Yang et~al\mbox{.}(2022)]%
        {yang2022getnext}
\bibfield{author}{\bibinfo{person}{Song Yang}, \bibinfo{person}{Jiamou Liu}, {and} \bibinfo{person}{Kaiqi Zhao}.} \bibinfo{year}{2022}\natexlab{}.
\newblock \showarticletitle{GETNext: trajectory flow map enhanced transformer for next POI recommendation}. In \bibinfo{booktitle}{\emph{Proceedings of the 45th International ACM SIGIR Conference on research and development in information retrieval}}. \bibinfo{pages}{1144--1153}.
\newblock


\bibitem[Yin et~al\mbox{.}(2023)]%
        {Shu2023MLLM}
\bibfield{author}{\bibinfo{person}{Shukang Yin}, \bibinfo{person}{Chaoyou Fu}, \bibinfo{person}{Sirui Zhao}, \bibinfo{person}{Ke Li}, \bibinfo{person}{Xing Sun}, {and} \bibinfo{person}{Chen~Enhong Xu~Tong}.} \bibinfo{year}{2023}\natexlab{}.
\newblock \showarticletitle{A Survey on Multimodal Large Language Models}.
\newblock \bibinfo{journal}{\emph{arXiv preprint arXiv:2306.13549}} (\bibinfo{year}{2023}).
\newblock


\bibitem[Zeghidour et~al\mbox{.}(2021)]%
        {zeghidour2021soundstream}
\bibfield{author}{\bibinfo{person}{Neil Zeghidour}, \bibinfo{person}{Alejandro Luebs}, \bibinfo{person}{Ahmed Omran}, \bibinfo{person}{Jan Skoglund}, {and} \bibinfo{person}{Marco Tagliasacchi}.} \bibinfo{year}{2021}\natexlab{}.
\newblock \showarticletitle{Soundstream: An end-to-end neural audio codec}.
\newblock \bibinfo{journal}{\emph{IEEE/ACM Transactions on Audio, Speech, and Language Processing}}  \bibinfo{volume}{30} (\bibinfo{year}{2021}), \bibinfo{pages}{495--507}.
\newblock


\bibitem[Zhang et~al\mbox{.}(2022)]%
        {zhang2022next}
\bibfield{author}{\bibinfo{person}{Lu Zhang}, \bibinfo{person}{Zhu Sun}, \bibinfo{person}{Ziqing Wu}, \bibinfo{person}{Jie Zhang}, \bibinfo{person}{Yew~Soon Ong}, {and} \bibinfo{person}{Xinghua Qu}.} \bibinfo{year}{2022}\natexlab{}.
\newblock \showarticletitle{Next Point-of-Interest Recommendation with Inferring Multi-step Future Preferences.}. In \bibinfo{booktitle}{\emph{IJCAI}}. \bibinfo{pages}{3751--3757}.
\newblock


\bibitem[Zhao et~al\mbox{.}(2020)]%
        {zhao2020go}
\bibfield{author}{\bibinfo{person}{Pengpeng Zhao}, \bibinfo{person}{Anjing Luo}, \bibinfo{person}{Yanchi Liu}, \bibinfo{person}{Jiajie Xu}, \bibinfo{person}{Zhixu Li}, \bibinfo{person}{Fuzhen Zhuang}, \bibinfo{person}{Victor~S Sheng}, {and} \bibinfo{person}{Xiaofang Zhou}.} \bibinfo{year}{2020}\natexlab{}.
\newblock \showarticletitle{Where to go next: A spatio-temporal gated network for next poi recommendation}.
\newblock \bibinfo{journal}{\emph{IEEE Transactions on Knowledge and Data Engineering}} \bibinfo{volume}{34}, \bibinfo{number}{5} (\bibinfo{year}{2020}), \bibinfo{pages}{2512--2524}.
\newblock


\bibitem[Zheng et~al\mbox{.}(2024)]%
        {Zheng2024LCRec}
\bibfield{author}{\bibinfo{person}{Bowen Zheng}, \bibinfo{person}{Yupeng Hou}, \bibinfo{person}{Hongyu Lu}, \bibinfo{person}{Yu Chen}, \bibinfo{person}{Wayne~Xin Zhao}, \bibinfo{person}{Ming Chen}, {and} \bibinfo{person}{Ji-Rong Wen}.} \bibinfo{year}{2024}\natexlab{}.
\newblock \showarticletitle{Adapting Large Language Models by Integrating Collaborative Semantics for Recommendation}. In \bibinfo{booktitle}{\emph{IEEE 40th International Conference on Data Engineering (ICDE)}}. \bibinfo{pages}{1435--1448}.
\newblock


\end{thebibliography}
\clearpage
\appendix
\section{Appendix}
\label{appendix}

\subsection{Data Augment}
\label{A.1}
\subsubsection{Multiple cropping}
If the input sequence length exceeds several times the given length, the sequence is divided into multiple non-overlapping segments. If the input sequence length is greater than the given length but less than twice that length, it is randomly cropped into two overlapping segments. If the input sequence length is shorter than the given length, no cropping is performed.

\begin{figure}[h]
    \centering
    \includegraphics[width=0.75\linewidth]{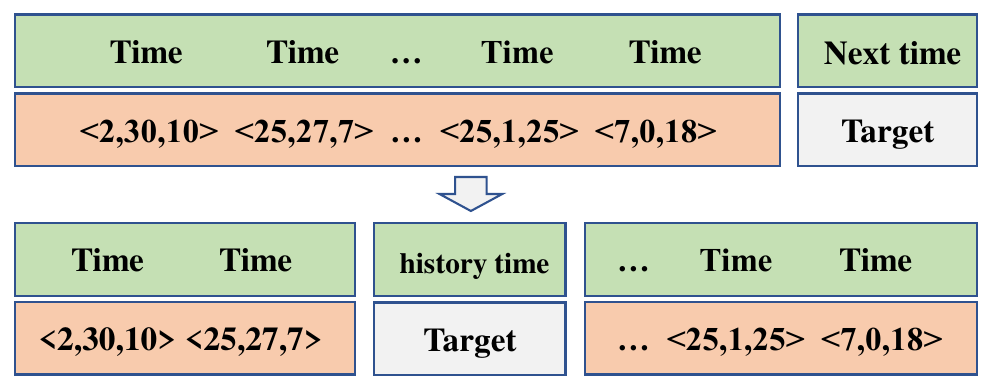}
    \caption{Replace the next POI recommendation task with a fill-in-the-blank task.}
    \label{fig:A1}
\end{figure}

\subsubsection{ Fill-in-the-blank}
For the cropped dataset, 1 out of every 5 samples is randomly selected to replace the next POI recommendation task with a fill-in-the-blank task, where the target is the POI corresponding to a randomly chosen timestamp from the historical check-in data.
The specific operation is shown in Figure~\ref{fig:A1}. 

\subsection{Baseline}
\label{A.2}
We introduce the details of the baseline methods we used in our experiments:
\begin{itemize}
    \item PRME~\cite{feng2015personalized}: A metric embedding-based method, which jointly incorporates sequential transition, user preference, and geographical influence.
    \item LSTM~\cite{graves2012long}: A variant of the RNN model, which handles long sequence dependency and Vanishing gradient problems.
    \item PLSPL~\cite{wu2020personalized}: A RNN-based method, which learns the user’s long-term pattern by using attention mechanism and shortterm preference. and short-term preference with LSTM.
    \item STAN~\cite{luo2021stan}: A transformer-based method, which modifies the attention coefficients using temporal and spatial interval information.
    \item GETNext~\cite{yang2022getnext}: A transformer and GNN-based method, which employs a user-agnostic global trajectory flow map and a novel Graph Enhanced Transformer model.
    \item STHGCN~\cite{yan2023spatio}: Constructing a hypergraph to capture inter and intra-user relations, STHGCN proposes a hypergraph transformer and solves the cold-start problem.
    \item TPG~\cite{luo2023timestamps}: A transformer-based method, which explicitly uses target times as prompts for the geography-aware location recommendation.
    \item ROTAN~\cite{feng2024rotan}: A time-aware next POI recommendation framework that incorporates target time information into the recommendation system. It introduces a novel time-aware attention mechanism based on rotation, utilizing the Time2Ro\-tation technique. This technique represents time intervals as rotational vectors and applies rotational operations to more effectively capture the influence of time information on user behavior.
    \item LLM4POI~\cite{li2024large}: The first LLM-based next POI recommendation method. By leveraging the commonsense knowledge of LLMs, the model can understand the underlying meaning of contextual information. Designing a specific prompt for POIs to preserve heterogeneous location-based social network data in its original format, hence avoiding the loss of contextual information. 
\end{itemize}

\subsection{Prompt format}
\label{A.3}
In Section \ref{fine-tune}, we provided the input format for the model under normal conditions. 
For the ablation experiments, the inputs for W/O SID and W/O time are shown as follows. 

\begin{table}[ht]
  \centering
  \caption{The input format of w/o SID}
  \renewcommand{\arraystretch}{1.1}
  \setlength{\tabcolsep}{3pt}       
  \begin{tabularx}{\columnwidth}{>{\bfseries}m{1.8cm} X}
    \hline
    Instruction & Here is a record of a user's POI accesses, your task is based on the history to predict the POI that the user is likely to access at the specified time. \\
    \hline
    Input & The user\_\textcolor{green}{<uid>} visited: \textcolor{blue}{<RID>} at \textcolor{orange}{[time]}, ..., visited \textcolor{blue}{<RID>} at \textcolor{orange}{[time]}. When \textcolor{red}{[time]} user\_\textcolor{green}{<uid>} is likely to visit:  \\
    \hline
    Example & The user\_<1>, visited <3312> at 2012-05-21 23:15, ..., visited <4345> at 2013-01-29 22:21, When 2013-01-31 14:05, user\_<1> is likely to visit: \\
    \hline
  \end{tabularx}
  \label{tab: w/o SID}
\end{table}

\begin{table}[ht]
  \centering
  \caption{The input format of w/o time}
  \renewcommand{\arraystretch}{1.1}
  \setlength{\tabcolsep}{3pt}       
  \begin{tabularx}{\columnwidth}{>{\bfseries}m{1.8cm} X}
    \hline
    Instruction & Here is a record of a user's POI accesses, your task is based on the history to predict the next POI that the user is likely to access. \\
    \hline
    Input & The user\_\textcolor{green}{<uid>} visited: \textcolor{blue}{<SID>}, ..., visited \textcolor{blue}{<SID>}, and in the next time user\_\textcolor{green}{<uid>} is likely to visit:  \\
    \hline
    Example & The user\_<1>, visited <a\_20> <b\_21> <c\_19> <d\_1>, <a\_11> <b\_19> <c\_6>  ..., <a\_11> <b\_19> <c\_6>, and in the next time user\_<1> is likely to visit:\\
    \hline
  \end{tabularx}
  \label{tab: w/o time}
\end{table}


\end{document}